\documentclass[twocolumn,superscriptaddress,reprint,nofootinbib,prd]{revtex4-1}%
\usepackage{amssymb}
\usepackage{color}
\usepackage{graphicx}
\usepackage{multirow}
\usepackage{dcolumn}
\usepackage{bm}
\usepackage[header,title,page,titletoc]{appendix}
\usepackage{amsmath}
\usepackage{bbm}
\usepackage{amsfonts}%
\usepackage[bookmarks,colorlinks=false]{hyperref}
\usepackage{ulem}
\usepackage{epstopdf}
\newlength{\wordlength}

\newlength{\onewordlength}

    \newcommand{\ba}{\begin{eqnarray}}
    \newcommand{\ea}{\end{eqnarray}}
    \newcommand{\be}{\begin{equation}}
    \newcommand{\ee}{\end{equation}}

\newcommand{\bzero}{{\bf 0}}
\newcommand {\btheta} {{\mbox{\boldmath$\theta$}}}

\newcommand{\bfe}{{\bf e}}

\newcommand {\bk} {{\mathbf k}}

\newcommand {\bp} {{\mathbf p}}

\newcommand{\bx}{{\bf x}}

\newcommand{\calP}{{\mathcal P}}

\newcommand{\calO}{{\mathcal O}}
\newcommand{\calR}{{\mathcal R}}

\newcommand{\calZ}{{\mathcal Z}}

\begin{document}

\title{Low-energy Scattering of $(D^{*}\bar{D}^{*})^\pm$ System and the Resonance-like Structure $Z_c(4025)$}

\author{Ying Chen}
\affiliation{%
Institute of High Energy Physics, Chinese Academy of Sciences, Beijing 100049, China
}%

\author{Ming Gong}
\affiliation{%
Institute of High Energy Physics, Chinese Academy of Sciences, Beijing 100049, China
}
\author{Yu-Hong Lei}
\affiliation{%
School of Physics, Peking University, Beijing 100871, China
}%

\author{Ning Li}
\affiliation{%
School of Science, Xi'an Technological University, Xi'An 710032, China
}%

\author{Jian Liang}
\affiliation{%
Institute of High Energy Physics, Chinese Academy of Sciences, Beijing 100049, China
}%

\author{Chuan Liu}%
\email[Corresponding author. Email: ]{liuchuan@pku.edu.cn}
\affiliation{%
School of Physics and Center for High Energy Physics, Peking
University, Beijing 100871, China
}%
\affiliation{Collaborative Innovation Center of Quantum Matter, Beijing 100871, China}


\author{Jin-Long Liu}
\affiliation{%
School of Physics, Peking University, Beijing 100871, China
}


\author{Yong-Fu Liu}
\affiliation{%
School of Physics, Peking University, Beijing 100871, China
}

\author{Yu-Bin Liu}
\affiliation{%
School of Physics, Nankai University, Tianjin 300071, China
}

\author{Zhaofeng Liu}
\affiliation{%
Institute of High Energy Physics, Chinese Academy of Sciences, Beijing 100049, China
}

\author{Jian-Ping Ma}
\affiliation{%
Institute of Theoretical Physics, Chinese Academy of Sciences, Beijing 100190, China
}

\author{Zhan-Lin Wang}
\affiliation{%
School of Physics, Peking University, Beijing 100871, China
}


\author{Jian-Bo~Zhang}
\affiliation{%
Department of Physics, Zhejiang University, Hangzhou 311027, China
}

 \begin{center}
 (CLQCD Collaboration)
 \end{center}
 \begin{abstract}
 In this paper, low-energy scattering of the
 $(D^{*}\bar{D}^{*})^\pm$ meson system is studied within L\"uscher's
 finite-size formalism using $N_{f}=2$ twisted mass gauge field
 configurations. With three different pion mass values,
 the $s$-wave threshold scattering parameters, namely the scattering
 length $a_0$ and the effective range $r_0$, are extracted in $J^P=1^+$ channel.
 Our results indicate that, in this particular channel,
 the interaction between the two vector charmed mesons is weakly repulsive
 in nature hence do not support the possibility of a shallow bound state for the
 two mesons, at least for the pion mass values being studied.  This study
 provides some useful information on the nature of the newly discovered
 resonance-like structure $Z_c(4025)$ observed in various experiments.
 \end{abstract}

 \maketitle



 \section{Introduction}
 Since the observation of charged charmonium-like structure
 $Z_{c}(3900)$~\cite{Ablikim:2013mio},
  the BESIII Collaboration studied the process
 $e^{+}e^{-} \rightarrow \pi^{\pm}(D^{*}\bar{D}^{*})^{\pm}$ at
 a center-of-mass energy of $4.26$~GeV and reported a new
 charged charmonium-like structure which they named as
 $Z^{\pm}_{c}(4025)$~\cite{Ablikim:2013emm},
 with a mass of 4026.3$\pm$2.6$\pm$3.7 MeV and a width of
 24.8$\pm$5.6$\pm$7.7 MeV. Such charged charmounium-like states
 are quite unique in the sense that their valence quark component
 must contain tetra-quark content $\bar{q}_1q_2\bar{c}c$ where $q_1$ and $q_2$ being
 two different flavors of light quark. Another feature is that, their mass values
 are rather close to the threshold of two corresponding charmed mesons.
 It is therefore tempting to explain these new exotic states as
 shallow bound states of the corresponding mesons. Another explanation
 is that they are simply genuine tetra-quark hadrons or mixture of
 the tetra-quark and the two-meson system.
 Since it is still unclear whether these states are above or below
 the threshold, it is also possible that they are resonances,
 or even simply cusp effects due to interaction between different
 channels. Obviously, a better understanding of the internal
 structures of these states will provide new insights into the
 dynamics of multi-quark systems and QCD low-energy behaviors.

 The experimental discovery of the charged charmonium-like structures have triggered
 a lot of theoretical studies in recent years, both using phenomenological
 methods~\cite{Chen:2013omd,He:2013nwa,Qiao:2013dda,Cui:2013xla} and on the lattice~\cite{Prelovsek:2014swa,Chen:2014afa}.
 Since $Z^{\pm}_{c}(4025)$ is near the $D^{*}\bar{D}^{*}$ threshold,
 a shallow bound state, also known as the molecular state, formed by $D^{*}$ and $\bar{D}^{*}$ mesons
 is a possible explanation. To further investigate
 this possible scenario, the interaction between $D^{*}$ and $\bar{D}^{*}$ mesons
 at low-energies becomes important. The energy considered here is very close to
 the threshold of the $D^{*}$-$\bar{D}^{*}$ system.  Therefore the
 interaction between the charmed mesons is non-perturbative
 in nature which requires a genuine non-perturbative framework such as lattice QCD.

 In this paper, the near-threshold scattering of
 $(D^{*}\bar{D}^{*})^\pm$ system is studied using $N_{f}=2$ twisted mass gauge field
 configurations~\cite{Blossier:2010cr}.
 The study is carried out for three different values of pion mass corresponding to
 $m_{\pi}=300,420,485$~MeV and the size of the lattices
 is $32^{3}\times 64$ with a lattice spacing of about $0.067$~fm.
 According to the BESIII results~\cite{Ablikim:2013emm}, the state $Z_c(4025)$ is
 consistent with the quantum number assignment $J^P=1^+$ although other assignments are
 not completely ruled out. Taking into the fact that the state is so close
 to the threshold where presumably $s$-wave scattering will dominate, we
 will focus on the $J^P=1^+$ channel only. Experiments also indicates that the state is
 strongly coupled to the $D^*\bar{D}^*$ system. Thus, in this exploratory lattice study,
 single-channel scattering of a $D^*$ and a $\bar{D}^*$ meson
 is studied using L\"uscher's formalism~\cite{Luscher:1990ux}.
 The $s$-wave low-energy scattering parameters, namely the scattering length $a_0$
 and the effective range $r_0$, are extracted from our simulation.
 To enhance the energy resolution close the threshold, twisted boundary
 conditions are utilized.

 This paper is organized as follows. In section~\ref{sec:theoretical-frame}, we briefly
 recapitulate L\"uscher's formalism in general and in the particular case of twisted boundary conditions.
 Section~\ref{sec:operators-and-correlators} defines the one-particle
 and two-particle interpolating operators used in this study and
 the corresponding correlation functions.
 In section~\ref{sec:simulation-details}, simulation
 details are provided and the results for the single-meson and two-meson systems
 are analyzed.  By applying L\"uscher's formula,
 the scattering phases are extracted and when fitted to the known low-energy behavior,
 the threshold scattering parameters of the system, i.e. the inverse scattering length $a^{-1}_0$
 and the effective range $r_0$ are obtained. As a crosscheck, both the jackknife and
 the bootstrap method have been used in this study which yield compatible results.
 Implications of our results are discussed afterwards.
 We finally conclude in section~\ref{sec:conclusions} with some general remarks.

 \section{Theoretical Framework}
 \label{sec:theoretical-frame}
 Let us first consider a particle with a mass $m$ enclosed in a cubic box of size $L\times L\times L$,
 then the ordinary periodic boundary condition in the spatial directions reads
\begin{equation}
\label{eq:PBC}
\Psi(\mathbf{x} + L\bfe_i, t ) = \Psi(\mathbf{x}, t)~,
\end{equation}
with the Cartesian unit vector $\bfe_i$ along the $i$-th axis ($i= 1,2,3$ for $x$, $y$, $z$ direction).
 The spatial momentum $\mathbf{k}$ of this particle is quantized according to:
\begin{equation}
\label{eq:free_k3}
\mathbf{k}=\frac{2\pi}{L}\mathbf{n},~~~\mathbf{n}\in\mathbb{Z}^{3}~.
\end{equation}
Now consider two interacting particles with masses $m_{1}$ and $m_{2}$ in this finite box. Taking
the center-of-mass frame of this system, the two particles thus have opposite three-momentum $\mathbf{k}$ and
$-\mathbf{k}$. The exact energy $E_{1.2}$ of the two-particle system is parameterized as
\begin{equation}
\label{eq:two particle system}
E_{1.2}(\mathbf{k})=\sqrt{m_{1}^{2}+\bar{\mathbf{k}}^2}+\sqrt{m_{2}^{2}+\bar{\mathbf{k}}^2}~,
\end{equation}
where $\bar{\mathbf{k}}^2$ is a quantity which also encodes the interaction
of the two particles in this box.
To be specific, $\bar{\mathbf{k}}^2=\mathbf{k}^{2}$ corresponds
to the non-interacting case, while $\bar{\mathbf{k}}^2 >\mathbf{k}^{2}$ and
$\bar{\mathbf{k}}^2 <\mathbf{k}^{2}$ corresponds
to repulsive or attractive interaction, respectively.
Based on Eq.~(\ref{eq:two particle system}), it is more convenient to define a
dimensionless quantity $q^2$:
\begin{equation}
\label{eq:q2 definition}
q^{2}=\frac{\bar{\mathbf{k}}^{2}L^{2}}{(2\pi)^2}~,
\end{equation}
such that the repulsive and attractive interactions are translated into
$q^{2}>\mathbf{n}^{2}$  and $q^{2}<\mathbf{n}^{2}$ for
some $\mathbf{n}\in\mathbb{Z}^{3}$, respectively.

In an actual lattice computation, the exact energy $E_{1.2}$ of the two-particle system
hence also the value of $q^2$ is obtained from corresponding correlation functions.
 L\"uscher's formula relates the value of $q^2$ and
 the elastic scattering phase shift $\delta(q)$ at that particular energy
 in the infinite volume.
 In the simplest case of $s$-wave elastic scattering, it reads:~\cite{Luscher:1990ux}
 \begin{equation}
 \label{eq:luscher_cube}
 q\cot\delta_0(q)={1\over \pi^{3/2}}\calZ_{00}(1;q^2)~,
 \end{equation}
 where $\calZ_{00}(1;q^2)$ is the zeta-function which
 can be evaluated numerically once its argument $q^2$ is given.
 Eq.~(\ref{eq:luscher_cube}) is the main formula to compute
 the elastic scattering phase shift on the lattice.
 In the case of attractive interaction,
 the lowest two-particle energy level can become lower than the threshold. If the interaction is weak,
 the state is loosely bound, i.e. $(-q^2)$ being positive
 but close to zero~\cite{Prelovsek:2013sxa,Sasaki:2007kr}. However, a negative $q^2$ value
 in a finite volume alone does not signifies a bound state. One has to
 investigate the behavior of the negative energy shift in the large volume limit.

 With quantization condition on three-momenta, c.f. Eq.~(\ref{eq:free_k3}),
the typical size of the smallest nonzero momentum is still too large to
investigate the hadron-hadron near-threshold scattering for practical size of the lattice.
We thus utilize the so-called twisted boundary conditions in our study~\cite{Bedaque:2004kc,Sachrajda:2004mi}.
 Following the notation in Ref.~\cite{Ozaki:2012ce}, the quark field $\psi_\btheta(\bx,t)$, when
 transported by an amount of $L$ along
 the spatial direction $i$ (designated by unit vector $\bfe_i$, $i=1,2,3$),
 will acquire an additional phase $e^{i\theta_i}$:
 \begin{equation}
 \label{eq:twistBC}
 \psi_{\btheta}(\bx+L\bfe_i,t)=e^{i\theta_i}\psi_{\btheta}(\bx,t)~,
 \end{equation}
 where $\btheta=(\theta_1,\theta_2,\theta_3)$ is the twisted angle (vector) for
 the quark field in three spatial directions.
 The conventional periodic boundary conditions corresponds to $\btheta=(0,0,0)$.
 Twisted boundary conditions such as those in Eq.~(\ref{eq:twistBC}) can be applied
 to any flavor of quark fields in question. In other words, we are free to choose a
 twisting angle vector $\btheta_f$ for flavor $f$, with $f=u,d,s,c,\cdots$.
 Under twisted boundary conditions, the discretized momentum
 in the finite volume is also modified. So, instead of Eq.~(\ref{eq:free_k3}), we have,
\begin{equation}
\label{eq:free_p3_TBC}
\mathbf{p} = \frac{2\pi}{L}\left(\mathbf{n} + \frac{\btheta}{2\pi}\right)~.
\end{equation}
 It is more convenient to introduce the new fields $\psi'$,
 we shall call them the primed fields, via
 \begin{equation}
 \psi'(\mathbf{x},t)=e^{-i\btheta.\bx/L}\psi_{\btheta}(\mathbf{x},t)~.
 \end{equation}
 It is easy to verify that the primed fields $\psi'(\mathbf{x},t)$ satisfy
 the usual periodic boundary conditions, c.f. Eq.~(\ref{eq:PBC}).
 For Wilson-type fermions, we can easily calculate the primed quark propagators,
 which are Wick contractions of the primed fields,
 using a modified set of gauge fields (the primed gauge fields),
 $U'_{x,\mu}=e^{i\theta_{\mu}a/L}U_{x,\mu}$ with $\theta_\mu=(0,\btheta)$~\cite{Ozaki:2012ce,Chen:2014afa}.

 Traditional meson interpolating operators are constructed using the primed fields as a local bilinears,
$\mathcal{O}_{\Gamma}(\mathbf{x},t)=\bar{\psi'}_f\Gamma\psi'_{f'}(\mathbf{x},t)$
, where $f$ and $f'$ denoting flavor indices and $\Gamma$ being a Dirac gamma matrix.
By summing over the spatial coordinate $\mathbf{x}$ with appropriate three-momentum $\bp$,
\be
\calO'_\Gamma(\bp,t)=\sum_\bx\bar{\psi'}_f\Gamma\psi'_{f'}(\mathbf{x},t)e^{-i\bp\cdot\bx}\;,
\ee
one sees that the above operator in fact corresponds to an operator
built using the un-primed fields with three momentum: $\mathbf{p}+(\btheta_{f'}-\btheta_{f})/L$.
Since it is free to choose any values of $\btheta_f$ and $\btheta_{f'}$,
an improved resolution is achieved in momentum space.

Note that we have adopted twisted boundary conditions for the valence quark fields. This
is referred to as the partial twisting.
Strictly speaking, the same twisted boundary condition should be applied
both to the valence and to the sea quark fields which is called full twisting. It has been
shown recently that, in some cases, partial twisting is equivalent to
full twisting~\cite{Agadjanov:2013kja}.
In other cases, however, the corrections due to partially twisted boundary conditions
 are shown to be exponentially suppressed if the size of the box is large~\cite{Sachrajda:2004mi}.
 We will assume that these corrections are indeed negligible.
 \footnote{This makes sense since L\"uscher's formalism also requires that
 exponentially suppressed corrections are negligible anyway.}
 In the following calculations, only the light quark fields ($u$ and $d$) will be twisted
 while the charm quark fields remain
 un-twisted.  This choice carefully avoids potential problems that might have arisen due to annihilation diagrams
 in this process as suggested in Ref.~\cite{Agadjanov:2013kja}.

 \section{OPERATORS AND CORRELATORS}
  \label{sec:operators-and-correlators}

 As usual, the energies of single-particle and two-particle systems are obtained
 from corresponding correlation functions which are measured in our Monte Carlo simulation.
 Since the newly discovered $Z_c(4025)$ state is observed in
 both $D^*\bar{D}^*$ and the $h_c\pi$ channel~\cite{Ablikim:2013emm}, its quantum number is likely
 to be $I^G(J^P)=1^+(1^+)$. The closeness of its mass to the $D^*\bar{D}^*$ threshold suggests that
 it might be a candidate for $D^*$-$\bar{D}^*$ bound state.
 In order to investigate the scattering relevant to this scenario on the lattice, we need to construct
 the $D^*\bar{D}^*$ two-particle interpolating operators with the right quantum number mentioned above.
 In practice, for the one-particle operators of $D^{*\pm}$ and $\bar{D}^{*0}$,
 conventional quark bilinear operators for vector mesons are utilized.
 The desired two-particle operators for system in the $I^G(J^P)=1^+(1^+)$ channel are
 discussed in the following.
 Due to the difference in the symmetries,
 the cases of twisted boundary conditions and non-twisted boundary condition have to be
 treated somewhat differently .

 \subsection{Operators in the non-twisted case}

 Let us first consider the non-twisted case.
 For a single vector charmed meson and its anti-particle, we utilize the following local interpolating
 fields in real space:
 \begin{equation}
 \label{eq:single_operators_defs}
 [D^{\ast +}]:\ \calP_{i}(\bx,t) = [\bar{c}\gamma_{i} d](\bx,t)~,
 \end{equation}
 \begin{equation}
 [D^{\ast -}]:\ \bar{\cal{P}}_{i}(\bx,t)=[\bar{d}\gamma_{i} c](\bx,t)=[\calP_{i}(\bx,t)]^\dagger~,
 \end{equation}
 In the above equation, we have also indicated the quark flavor content of the operator
 in front of the definition inside the square bracket.
 So, for example, the operator in Eq.~(\ref{eq:single_operators_defs})
 will create a $D^{\ast +}$ meson when acting on the QCD vacuum.
 A single-particle state with definite three-momentum $\bk$ is
 defined accordingly via usual Fourier transform~\cite{Meng:2009qt}:
 \begin{equation}
 \label{eq:P-operator-definition}
  \mathcal{P}_{i}(\mathbf{k},t)=\sum_\mathbf{x} \mathcal{P}_{i}(\mathbf{x},t)e^{-i \mathbf{k} \cdot \mathbf{x}}~.
 \end{equation}
 The conjugate of the above operator is:
 \begin{equation}
 [\mathcal{P}_{i}(\mathbf{k},t)]^{\dag}=\sum_{\mathbf{x}} [\mathcal{P}_{i}(\mathbf{x},t)]^{\dag}
  e^{+i\mathbf{k}\cdot\mathbf{x}} \equiv \bar{\mathcal{P}}_{i}(-\mathbf{k},t)~.
 \end{equation}
 Similarly, for $\bar{D}^{\ast 0}$ and its anti-particle, we use the following operators:
 \begin{equation}
 \label{eq:Q-operator-definition}
 \begin{split}
 &[\bar{D}^{\ast 0}]: ~\mathcal{Q}_{i}(\bx,t) = [\bar{c}\gamma_{i} u](\bx,t)~,\\
 &[D^{\ast 0}]: ~\bar{\mathcal{Q}}_{i}(\bx,t)=[\bar{u}\gamma_{i} c](\bx,t)=[\mathcal{Q}_{i}(\bx,t)]^{\dagger}~,\\
 &\mathcal{Q}_{i}(\mathbf{k},t)=\sum_\mathbf{x} \mathcal{Q}_{i}(\mathbf{x},t)e^{-i \mathbf{k} \cdot \mathbf{x}}~,\\
 &[\mathcal{Q}_{i}(\mathbf{k},t)]^{\dag}=\sum_{\mathbf{x}} [\mathcal{Q}_{i}(\mathbf{x},t)]^{\dag}
  e^{+i\mathbf{k}\cdot\mathbf{x}} \equiv \bar{\mathcal{Q}}_{i}(-\mathbf{k},t)~.
 \end{split}
 \end{equation}

 For the two-particle operators, in terms of the operators defined above,
 we have used the following combination for a pair
 of charmed mesons with back-to-back momentum,
 \begin{equation}
 \mathcal{P}_{i}(\mathbf{k},t)\mathcal{Q}_{j}(-\mathbf{k},t)
 -\mathcal{P}_{j}(\mathbf{k},t)\mathcal{Q}_{i}(-\mathbf{k},t)~,
 \end{equation}
 with $i,j=1,2,3$.
 On a finite lattice, however, the rotational group $SO(3)$ is broken down
 to the cubic group $O_h$ and $J^P=1^+$ of the two-particle system is
 thus reduced to $T^+_1$ of the cubic group.
 To avoid complicated Fierz rearrangement terms, we have put the
 two mesons on two neighboring time-slices. Thus, we use the following
 operators to create the state with two charmed mesons,
\begin{equation}
\label{eq:T1-operator}
\mathcal{O}^{T_{1}}
\begin{cases}
\mathcal{O}^{T_{1}}_{1}:&\sum\limits_{R\in G}
             [\mathcal{P}_{2}(R\circ \mathbf{k}_{\alpha},t+1)\mathcal{Q}_{3}(-R\circ \mathbf{k}_{\alpha},t)\\
            &-\mathcal{P}_{3}(R\circ \mathbf{k}_{\alpha},t+1)\mathcal{Q}_{2}(-R\circ \mathbf{k}_{\alpha},t)]~,\\

\mathcal{O}^{T_{1}}_{2}:&\sum\limits_{R\in G}
            [\mathcal{P}_{1}(R\circ \mathbf{k}_{\alpha},t+1)\mathcal{Q}_{3}(-R\circ \mathbf{k}_{\alpha},t)\\
            &-\mathcal{P}_{3}(R\circ \mathbf{k}_{\alpha},t+1)\mathcal{Q}_{1}(-R\circ \mathbf{k}_{\alpha},t)] ~,\\

\mathcal{O}^{T_{1}}_{3}:&\sum\limits_{R\in G}
            [\mathcal{P}_{1}(R\circ \mathbf{k}_{\alpha},t+1)\mathcal{Q}_{2}(-R\circ \mathbf{k}_{\alpha},t)\\
            &-\mathcal{P}_{2}(R\circ \mathbf{k}_{\alpha},t+1)\mathcal{Q}_{1}(-R\circ \mathbf{k}_{\alpha},t)]~,

\end{cases}
\end{equation}
 where $\bk_\alpha$ is a chosen three-momentum mode. The index $\alpha$
 ($\alpha=1,\cdots,N$) denotes the momentum mode
 considered in our calculation. In this particular case, we have $N=4$.
 In the above equation, $G=O_{h}$ designates the cubic group
 and $R\in G$ is an element of the group and we have used the
 notation $R\circ\bk_\alpha$ to denote the momentum obtained from
 $\bk_\alpha$ by applying the operation $R$ on $\bk_\alpha$.

 Note that in the above constructions, we have not included relative orbital
 angular momentum of the two particles, i.e. we are only studying
 the $s$-wave scattering of the two mesons. This is justified for this
 particular case since close to the threshold, the scattering is always
 dominated by the $s$-wave contributions.

 \subsection{Operators in the case of twisted boundary conditions}

 As explained at the end of previous section,
 we choose to apply twisted boundary conditions to the light quarks ($u$ and $d$) while
 the charm quark remains un-twisted. Single meson operators are the same as
 in the previous subsection except that all the operators are
 constructed using the primed fields.
 We also set the twisting angle for the $u$ and $d$ quark fields
 to be identical so that their lattice propagators are related to each other
 by a simple conjugation in the twisted mass formalism.

 For the two-particle operators, the only difference is the discrete version of the rotational symmetry.
 It has been reduced from
 $O_h$ to one of its subgroups: $C_{4v}$, $D_{4h}$, $D_{2h}$, or $D_{3d}$, depending on
 the particular choice of $\btheta$.
 The other structures (flavor, parity when applicable etc.) of the operators remain unchanged.
 As a consequence, the operators $\mathcal{P}_{i}$ and $\mathcal{Q}_{i}$, which used to form a basis
 for the $T_1$ irrep of $O_h$ now have to be decomposed into new
 basis of the corresponding subgroups~\cite{Ozaki:2012ce,Chen:2013omd}:
 \begin{equation}
 \label{eq:reduction}
 \begin{array}{ll}
 T_1 \mapsto A_1 \oplus E                & ~~C_{4v}~,\\
 T_1 \mapsto A_2 \oplus E                & ~~D_{4h}~,\\
 T_1 \mapsto B_1 \oplus B_2 \oplus B_3   & ~~D_{2h}~,\\
 T_1 \mapsto A_2 \oplus E                & ~~D_{3d}~.
 \end{array}
 \end{equation}
 The information for these decompositions are summarized in Table~\ref{tab:twisted-symmetries}.
 As an example, take the first line of Eq.~(\ref{eq:reduction}) which corresponds to
 the case of $\btheta=(0,0,\pi)$,
 the original operator triplet $(\mathcal{P}_{1}$,$\mathcal{P}_{2}$,$\mathcal{P}_{3})$
 should be decomposed into a singlet $(\mathcal{P}_{3})$ and
 a doublet $(\mathcal{P}_{1}$,$\mathcal{P}_{2})$
 which forms the basis for $A_1$ and $E$ irreps, respectively.
 Similar relations also hold for the $\mathcal{Q}_{i}$'s.
 \footnote{The reason that $\mathcal{P}_{3}$ is special as
 opposed to $\mathcal{P}_{1}$ and $\mathcal{P}_{2}$ is
 because the twisted boundary condition with $\btheta=(0,0,\pi)$ is applied in
 the $3$-direction which breaks the symmetry.}

 The construction of the two-particle operators in the case of twisted boundary conditions
 is somewhat complex. Let us start from a general problem in group theory. Suppose that $e_{i}$ form
 the basis of a 3-dimensional irreps $T_{1}$ while $e'_{i}$ form the basis of another
 3-dimensional irreps $T_{1}$. With the help of group theory, the direct product of these two
 3-dimensional irreps can form a 9-dimensional reducible representation of basis
 $e_{i}\otimes e'_{j}(i,j=1,2,3)$. Depending on the particular choice of $\btheta$, this new
 9-dimensional reducible representation will be decomposed into irreps of the corresponding
 subgroup, with the linear combinations of $e_{i}\otimes e'_{j}(i,j=1,2,3)$
 giving the basis of these irreps.

 To find the linear combination of basis $e_{i}\otimes e'_{j}(i,j=1,2,3)$ for definite irrep
 we are interested in, one could use different approaches.
 In our study, group character technique is used to determine the specific basis for a certain irrep.


 As an application of this technique described above,
 taking the case of $\btheta=(\pi,\pi,0)$ as an example,
 we give the corresponding operators as listed in the following equations.
 \begin{equation}
 \begin{split}
 &B_{1}:~e_{1}\otimes e'_{2}+e_{2}\otimes e'_{1}~,\\
 &B_{2}:~e_{1}\otimes e'_{3}+e_{3}\otimes e'_{1}~,\\
 &B_{3}:~e_{2}\otimes e'_{3}+e_{3}\otimes e'_{2}~,\\
 \end{split}
 \end{equation}
where $e_{1}=\frac{1}{\sqrt{2}}(\mathcal{P}_{1}+\mathcal{P}_{2})$, $e_{2}=\frac{1}{\sqrt{2}}(\mathcal{P}_{2}-\mathcal{P}_{1})$, $e_{3}=\mathcal{P}_{3}$. Similar relations
 also hold between $e'_{i}$ and $\mathcal{Q}_{i}$. Then we have two-particle operators for irrep $B_{1}$
 as shown below:
 \begin{equation}
 \begin{split}
 \mathcal{O}^{B_{1}}:&\sum\limits_{R\in G}
            [\mathcal{P}_{2}(R\circ \mathbf{k}_{\alpha},t+1)\mathcal{Q}_{2}(-R\circ \mathbf{k}_{\alpha},t)\\
            &-\mathcal{P}_{1}(R\circ \mathbf{k}_{\alpha},t+1)\mathcal{Q}_{1}(-R\circ \mathbf{k}_{\alpha},t)]~.
 \end{split}
\end{equation}
 where $G=D_{2h}$, the group corresponding to $\btheta=(\pi,\pi,0)$.

\subsection{Correlation functions}

 For vector charmed meson $D^{*}$ and $\bar{D}^{*}$, the corresponding
 correlation functions are defined as:
 \begin{equation}
 \label{eq:one-particle-lattice}
 \begin{split}
 &C^{\mathcal{P}}(\mathbf{k},t)=\langle
       \mathcal{P}_{i}^{\dagger}(\mathbf{k},t)\mathcal{P}_{i}(\mathbf{k},0)\rangle~,\\
 &C^{\mathcal{Q}}(\mathbf{k},t)=\langle
       \mathcal{Q}_{i}^{\dagger}(\mathbf{k},t)\mathcal{Q}_{i}(\mathbf{k},0)\rangle~,
  \end{split}
 \end{equation}
 where $\mathbf{k}$ represents the three-momentum of the relevant particle.
 It is straightforward to obtain the single particle energy $E(\mathbf{k})$ for various lattice
 momentum $\mathbf{k}$. For the single particle, the dispersion relation can then be checked with
 various  $E(\mathbf{k})$. In particular, this can be checked in both twisted boundary conditions
 and conventional periodic boundary conditions. With judicious choices of $\btheta$, one could
 check the single-particle dispersion relation to a much better accuracy which will be
 shown in the next section.

 Two-particle correlation functions are somewhat more involved.
 Generally speaking, a correlation
 matrix $C^{\Gamma}_{\alpha\beta}(t)$ is constructed:
 \begin{equation}
 \label{eq:correlation-matrix}
 C^{\Gamma}_{\alpha\beta}(t)=\langle\mathcal{O}^{\Gamma\dagger}_{\alpha}(t)
                             \mathcal{O}^{\Gamma}_{\beta}(0)\rangle~.
 \end{equation}
 where $\mathcal{O}^{\Gamma}_{\alpha}$ represents the two-particle operator defined
 in the previous section and $\Gamma$ denotes a definite irrep while $\alpha$ enumerates
 different operators in that irrep. To be specific, for the non-twisted case $\btheta=(0, 0, 0)$,
 the number of $\bk_{\alpha}$ is 4 in $T_{1}$ channel while for all other cases,
 the number of $\bk_{\alpha}$ is 2. As a reference, these information
 are also collected in Table~\ref{tab:twisted-symmetries}.
\begin{table}[!htb]
\centering \caption{Information about the two-particle operators used in
 this calculation together with the corresponding symmetries.}
 \label{tab:twisted-symmetries}
\begin{tabular}{|c|c|c|c|c|c|}
\hline
$\btheta$    & ~$\bzero$ ~   & ~$(0,0,\frac{\pi}{2})$ ~ &~$(0,0,\pi)$ ~& ~$(\pi,\pi,0)$ ~
& ~$(\pi,\pi,\pi)$ \\
\hline
Symmetry  & $O_h$     & $C_{4v}$   & $D_{4h}$  & $D_{2h}$  & $D_{3d}$    \\
\hline
irreps & $T_1$  & $A_1$, $E$ &  $E$ & $B_1$, $B_2$, $B_3$ & $E$  \\
\hline
Number of $\bk_\alpha$     &4   &2~,~2  & 2  &2~,~2~,~2  & 2   \\
\hline
\end{tabular}
\end{table}

 \section{Simulation details and results}
 \label{sec:simulation-details}

 In this paper, the Osterwalder-Seiler action~\cite{Frezzotti:2004wz} is used for the valence charm quark.
 The gauge field ensemble comes from $N_f=2$ twisted mass gauge field configurations
 generated by the European Twisted Mass Collaboration (ETMC)~\cite{Blossier:2010cr}.
 The gauge coupling is $\beta=4.05$ which corresponds to a lattice spacing of about $0.067$~fm
 and we have used three different pion mass values, namely 300~MeV, 420~MeV and 485~MeV.
 Details of the relevant parameters are summarized in the Table~\ref{tab:parameter}.
 The up and down bare quark mass values, characterized by the bare quark parameter $\mu$
 in Table~\ref{tab:parameter}, are fixed to that of the sea-quark.
 For the charm quark, the mass parameter $a\mu_{c}$ is fixed so that
 the value of $\frac{1}{4}m_{\eta_{c}}+\frac{3}{4}m_{J/\Psi}$ calculated on
 the lattice reproduces the corresponding experimental value.
\begin{table}[!htb]
\centering \caption{Simulation parameters in this study.} \label{tab:parameter}
\begin{tabular}{|c|c|c|c|c|c|c|}
\hline
$~~~\mu~~~$      &$N_{\rm conf}$   &$m_\pi$[MeV]  &$~m_\pi L~$  &$~~L^3\times T~~$  &~~a[fm]~~  &$~~\beta~~$\\
\hline
$0.003$    &200              &300           &3.3        &$32^{3}\times 64$  &0.067   &4.05\\
\hline
$0.006$    &200              &420           &4.6        &$32^{3}\times 64$  &0.067   &4.05\\
\hline
$0.008$    &200              &485           &5.3        &$32^{3}\times 64$  &0.067   &4.05\\
\hline
\end{tabular}
\end{table}

  \subsection{One-particle spectrum and dispersion relation}
  \label{subsec:one-particle-dispersion}

\begin{figure}[!htb]
   {\resizebox{0.5\textwidth}{!}{\includegraphics{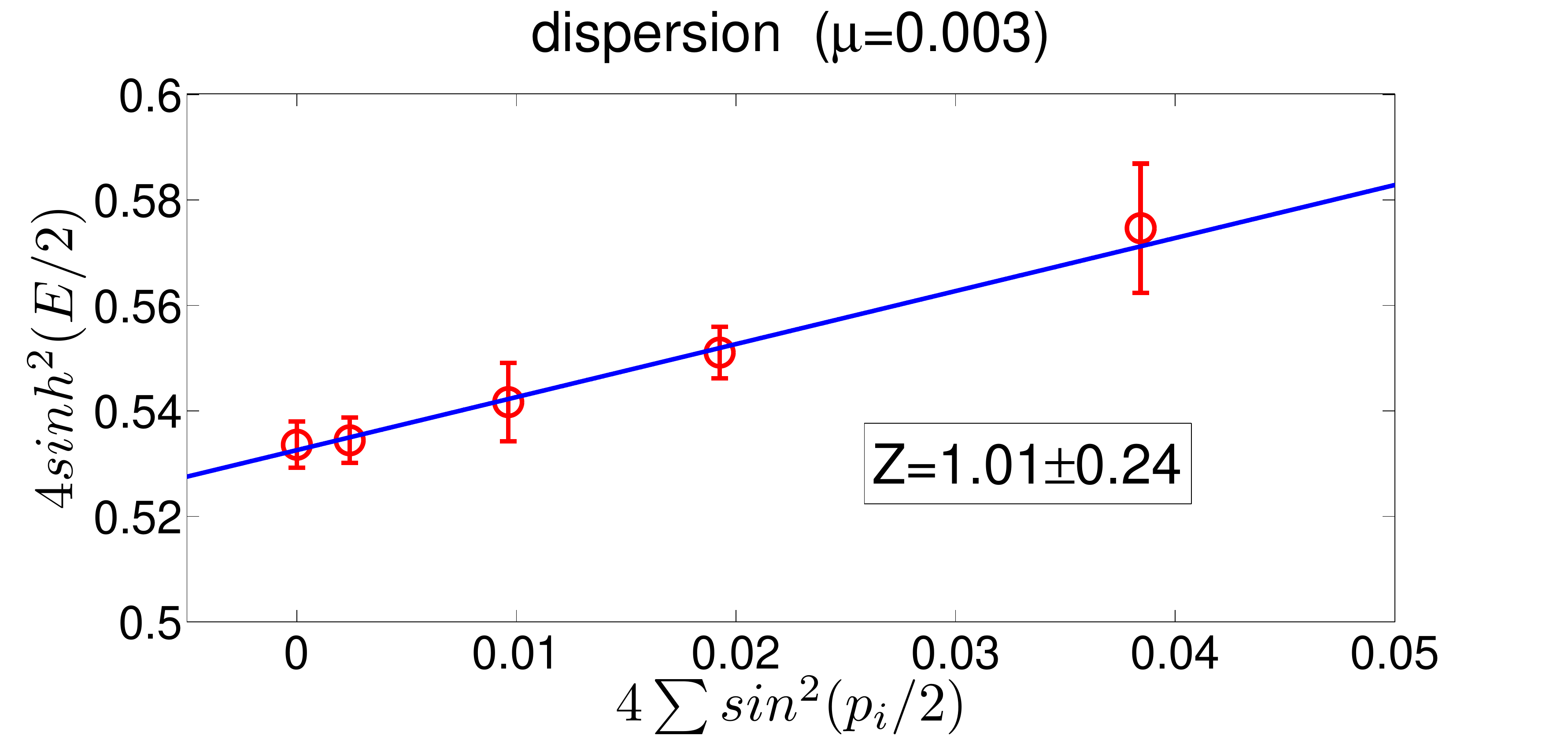}}}
   {\resizebox{0.5\textwidth}{!}{\includegraphics{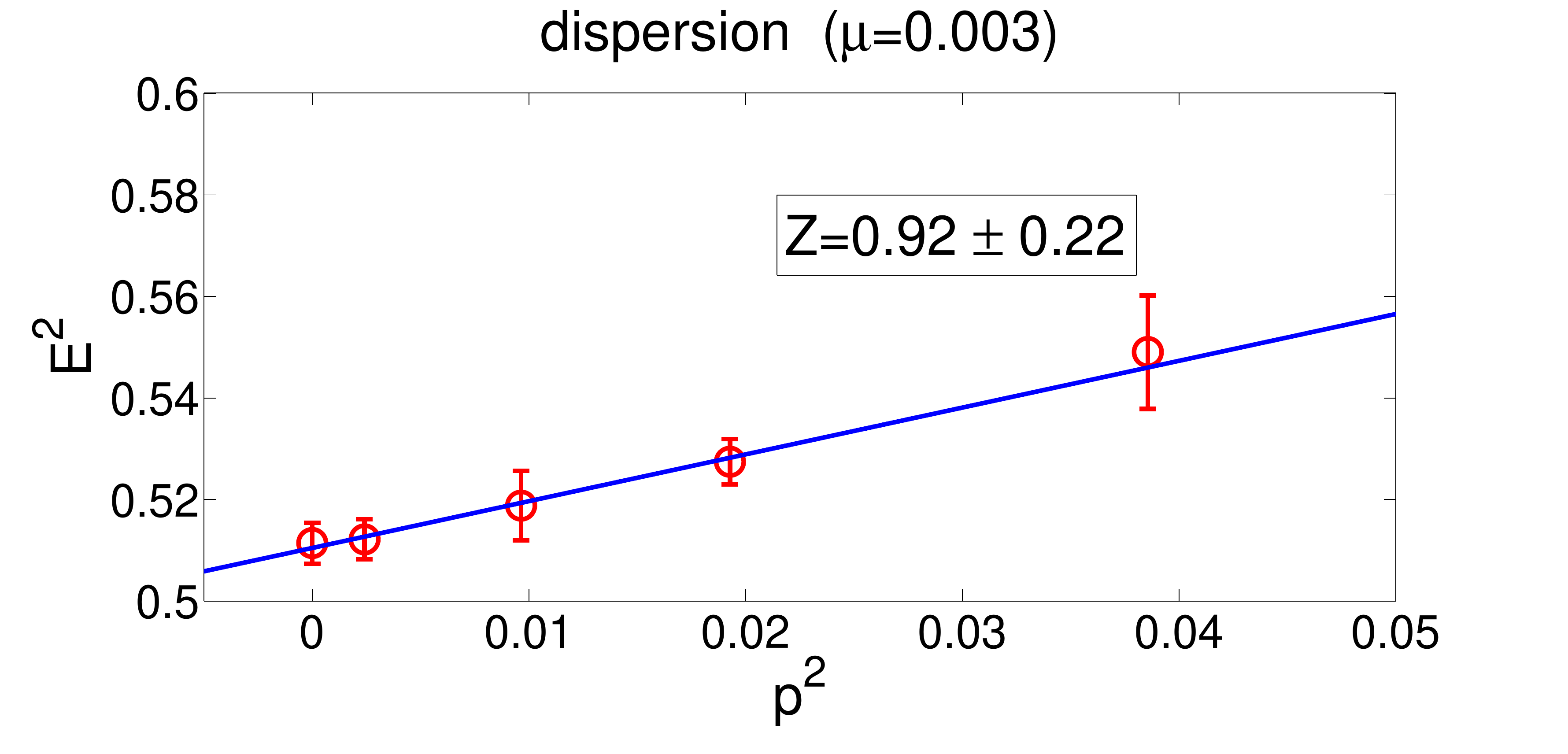}}}
  \caption{Dispersion relation for the $D^{*}$ meson at $\mu=0.003$ with
  lattice case (upper panel) and continuous case (lower panel).
  The points with error bars are lattice data while the straight lines are the
  corresponding linear fits with $Z$ denoting the slope of the line.
  \label{fig:single_particle_dipersions}}%
  \end{figure}

 One-particle correlation functions as defined in Eq.~(\ref{eq:one-particle-lattice}) with definite three-momentum $\bk$ are calculated in our simulation from which the one-particle
  spectrum $E(\bk)$ is obtained.
  We have checked the single particle dispersion relations for $D^{*}$ and $\bar{D}^{*}$ mesons,
  with both periodic boundary conditions and twisted boundary conditions. For the twisted
  boundary conditions, equivalent small momentum points offer us a more stringent test for
  the dispersion relation, both the continuum one and its lattice counterpart, at low-momenta
  close to zero. One example of these is illustrated in Fig.~\ref{fig:single_particle_dipersions}
  at $\mu=0.003$. The quantity $E(\bk)^2$ or its lattice counterpart
  $4\sinh^2(E/2)$ is shown versus $p^2$ or $\hat{\bp}^2=4\sum_i\sin^2(p_i/2)$
  in the bottom/top panel, respectively. The straight lines are linear fits
  with $Z$ being the fitted slope of the lines.

\subsection{Extraction of two-particle energy levels}
\label{subsec:two-particle-energies}

 In this paper, the usual L\"uscher-Wolff method~\cite{Luscher:1990ck}
 is adopted to extract the two-particle energy eigenvalues.
 For this purpose,  a new matrix $\Omega(t,t_0)$ is constructed as:
 \begin{eqnarray}
  \Omega(t,t_0)=C(t_0)^{-{1\over2}}C(t)C(t_0)^{-{1\over 2}}~,
 \end{eqnarray}
 where $t_0$ is a reference time-slice. Normally
 $t_0$ is picked such that the signal is good and stable.
 The energy eigenvalues for the two-particle system are
 then obtained by diagonalizing the matrix $\Omega(t,t_0)$.
 The eigenvalues of the matrix, $\lambda_\alpha(t,t_0)$, have the usual exponential decay behavior
 as described by $\lambda_\alpha\sim e^{-E_\alpha(t-t_0)}$ and therefore
 the exact energy $E_\alpha$ can be extracted from
 the effective mass plateau of the eigenvalue $\lambda_\alpha$.

 The real signal for the eigenvalue in our simulation turns out
 to be somewhat noisy. To enhance the signal, the following ratio was attempted:
  \begin{equation}
  \begin{split}
  \calR_\alpha(t,t_0)&={\lambda_\alpha(t,t_0)\over C^{\mathcal{P}}(t-t_0,\bzero)C^{\mathcal{Q}}(t-t_0,\bzero)}\\
                     &\propto e^{-\Delta E_\alpha\cdot (t-t_0)}~,
  \end{split}
  \end{equation}
 where $C^{\mathcal{P}}(t-t_0,\bzero)$ and $C^{\mathcal{Q}}(t-t_0,\bzero)$ are one-particle correlation
 functions with zero momentum for the corresponding mesons defined
 in Eq.~(\ref{eq:P-operator-definition}) and Eq.~(\ref{eq:Q-operator-definition}).
 Therefore, $\Delta E_\alpha$ is the difference of the two-particle
 energy measured from the threshold of the two mesons:
 \begin{equation}
 \label{eq:DeltaE_def}
  \Delta E_\alpha=E_\alpha-m_{D^\ast}-m_{\bar{D}^{\ast}}~.
 \end{equation}
 The energy difference $\Delta E_\alpha$ can
 be extracted from the plateau behavior of the effective mass function
 $\Delta E_{\alpha,{\rm eff}}(t)$ constructed from the
 ratio $\calR_\alpha(t,t_0)$ as usual:
 \be
 \Delta E_{\alpha,{\rm eff}}(t)=\ln\left({\calR_\alpha(t,t_0)\over \calR_\alpha(t+1,t_0)}\right)\;.
 \ee

\begin{figure}[htb]
   {\resizebox{0.5\textwidth}{!}{\includegraphics{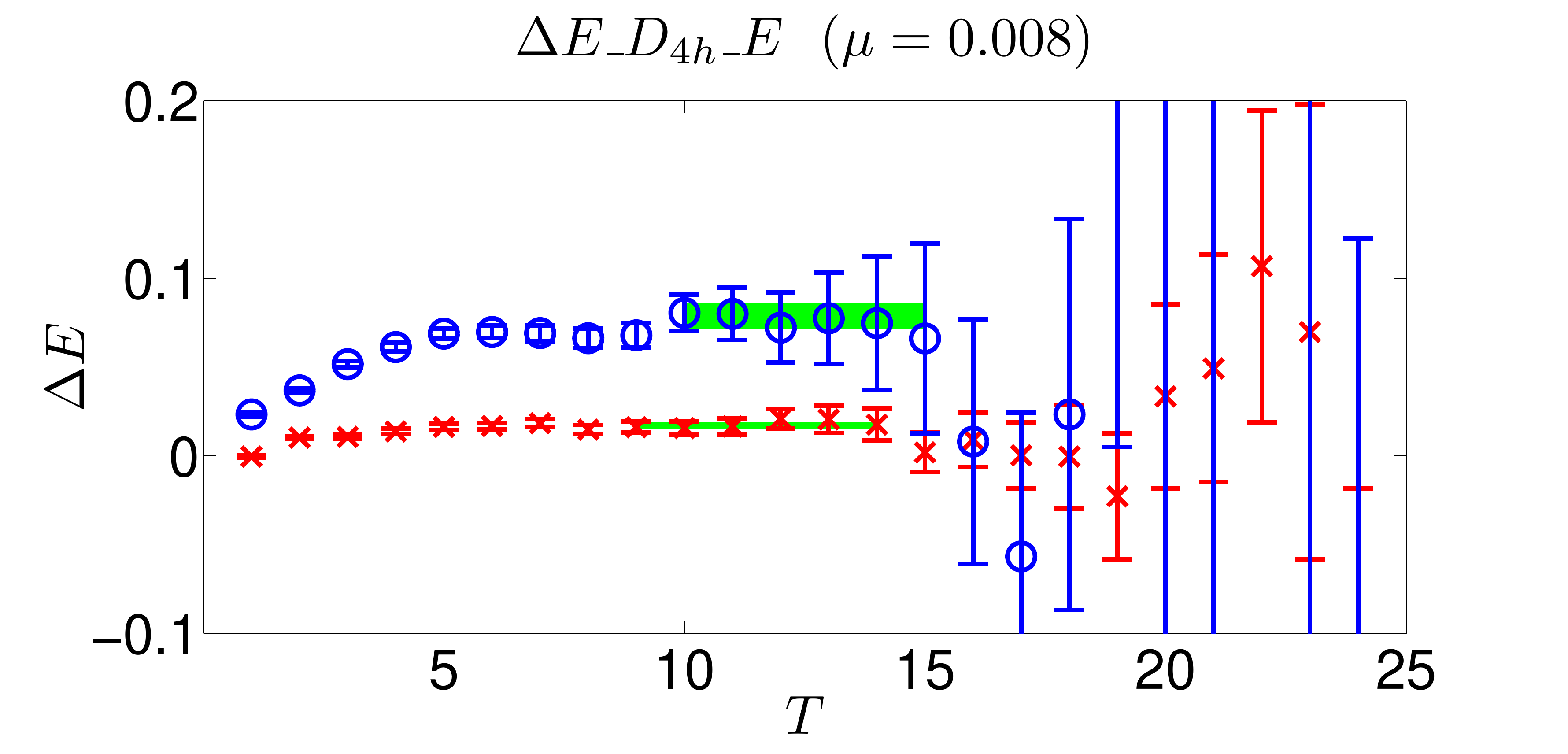}}}
   {\resizebox{0.5\textwidth}{!}{\includegraphics{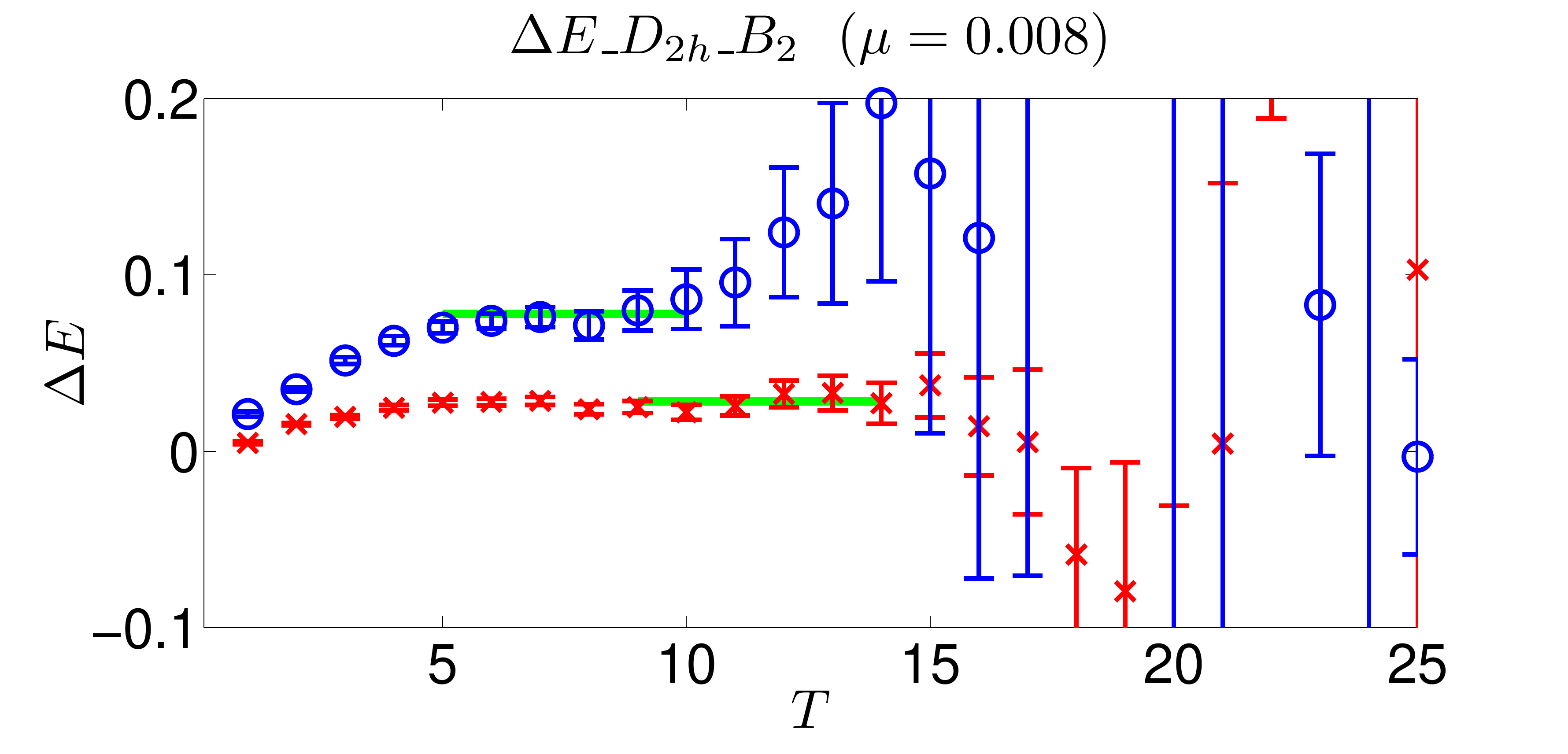}}}
 \caption{Effective mass plots for the energy shift $\Delta E_\alpha$ using the jackknife method
 at $\mu=0.008$ in the $E$ channel for $\btheta=(0,0,\pi)$ (top) and $B_{2}$ channel for
 $\btheta=(\pi,\pi,0)$ (bottom).
 The red crosses and the blue open circles correspond to two different
 energy levels obtained from Eq.~(\ref{eq:correlation-matrix})
 using $N=2$ different two-particle operators as discussed in the text.
  The horizontal bands indicate
  the fitted values for $\Delta E_\alpha$ and the
  corresponding fitting ranges.
  \label{fig:two-particle-plateaus-jackknife}}%
  \end{figure}

\begin{figure}[htb]
   {\resizebox{0.5\textwidth}{!}{\includegraphics{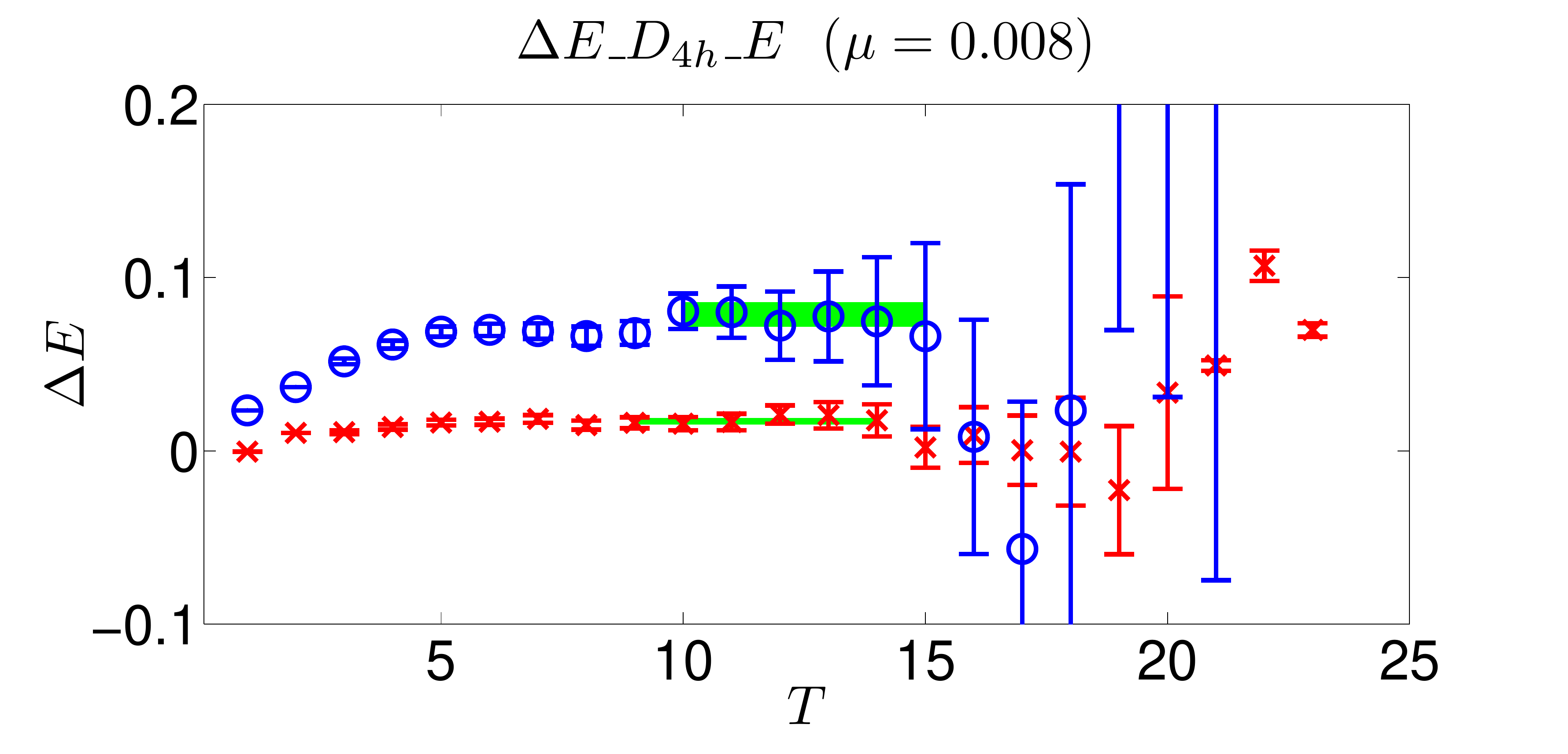}}}
   {\resizebox{0.5\textwidth}{!}{\includegraphics{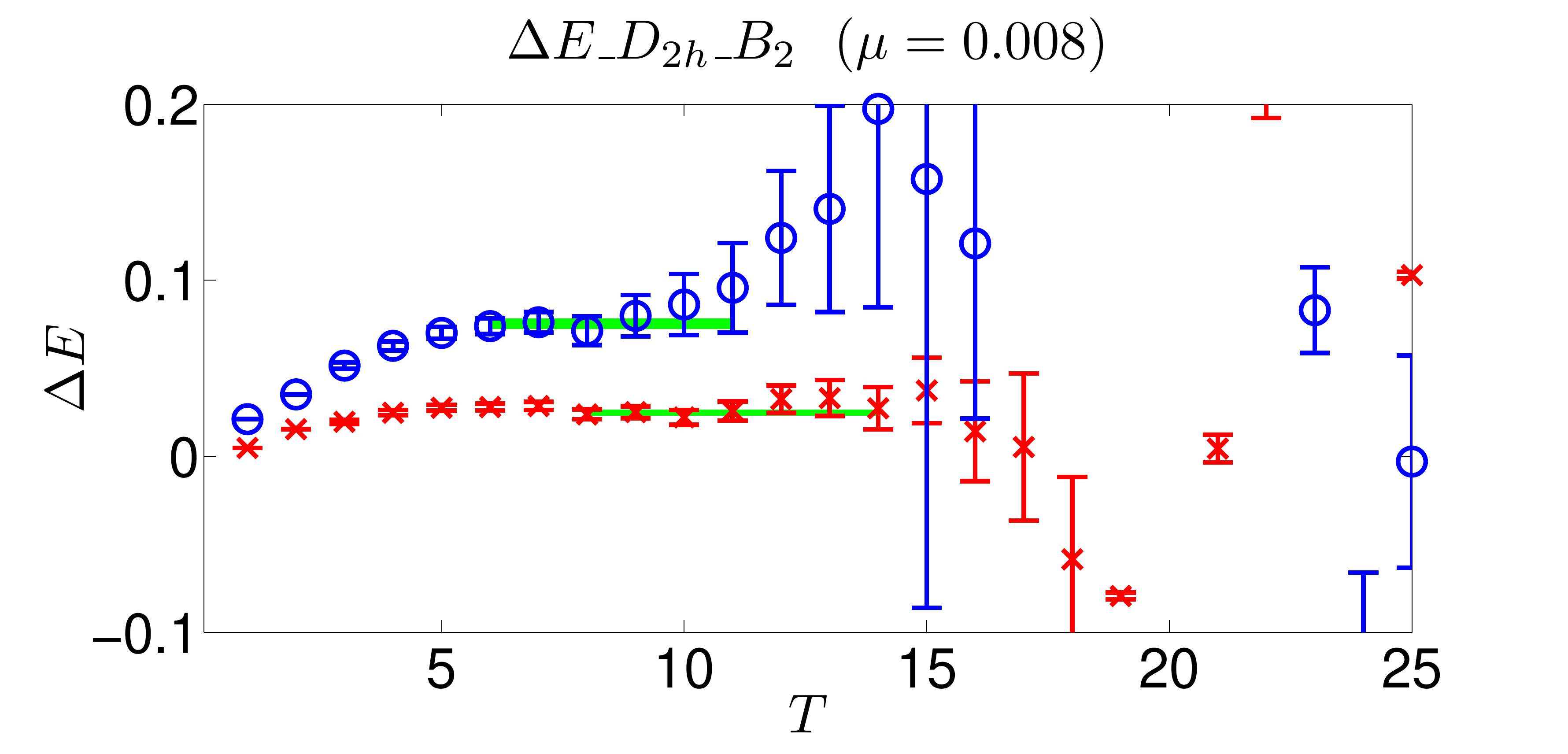}}}
\caption{Same as  Fig.~\ref{fig:two-particle-plateaus-jackknife} but using the bootstrap method.
  \label{fig:two-particle-plateaus-bootstrap}}%
  \end{figure}

 With the energy effective energy difference $\Delta E_{\alpha,{\rm eff}}(t)$ for each time slice $t$,
 we estimate the error for each $\Delta E_{\alpha,{\rm eff}}(t)$ using the jackknife method.
 Then, from the effective energy difference $\Delta E_{\alpha,{\rm eff}}(t)$ and its
 corresponding errors, one searches a plateau in $t$ that extends several consecutive time-slices
 and minimizes the $\chi^2$ per degree of freedom.
 From this procedure, a fitted value of $\Delta E_\alpha$ together with its error
 is obtained. As an illustration, in Fig.~\ref{fig:two-particle-plateaus-jackknife},
 we have shown the fitted values of $\Delta E_\alpha$ using the jackknife method at
 $\mu=0.008$ in the $E$ channel for $\btheta=(0,0,\pi)$, and
 the $B_{2}$ channel for $\btheta=(\pi,\pi,0)$.
 As a cross check, bootstrap method
 is also tried to calculate the standard error
 of $\Delta E_{\alpha,{\rm eff}}(t)$ on each time slice.
 To make the comparison, the fitted values of $\Delta E_\alpha$ are also illustrated in
 Fig.~\ref{fig:two-particle-plateaus-bootstrap} for the same cases as in
 Fig.~\ref{fig:two-particle-plateaus-jackknife}.

 It is seen graphically from Fig.~\ref{fig:two-particle-plateaus-jackknife} and
 Fig.~\ref{fig:two-particle-plateaus-bootstrap},
 the fitted values of $\Delta E_\alpha$ from jackknife method is consistent with those from
 bootstrap method within the statistical uncertainties.
 The only difference is that bootstrap method seems to give a somewhat smaller error
 of $\Delta E_{\alpha}(t)$ on each time slice. To be on the safe side, in this paper
 we regard the results from jackknife method as our final results for the energy levels.

  Effective mass plots for other cases are similar
  to those shown in Fig.~\ref{fig:two-particle-plateaus-jackknife} and
 Fig.~\ref{fig:two-particle-plateaus-bootstrap}.
 With the energy difference $\Delta E_\alpha$ extracted from the simulation data,
 one utilizes the definition:
 \begin{equation}
   \label{eq:kbar}
  \sqrt{m_{D^\ast}^2+\bar{\bf k}^2}+\sqrt{m_{\bar{D}^{\ast}}^2+\bar{\bf k}^2}
  =\Delta E_\alpha + m_{D^{\ast}} + m_{\bar{D}^{\ast}}~.
 \end{equation}
 to solve for $\bar{\bk}^2\equiv (2\pi/L)^2q^2$ which is then plugged into L\"uscher's formula
 to obtain the information about the scattering phase shift.

 The final results for $\Delta E_\alpha$ in each irrep, together with the corresponding ranges from which the
 $\Delta E_\alpha$'s are extracted, are summarized in Table~\ref{tab:deltaE}.
 We only keep the two lowest energy levels for the non-twisted case and the lowest for the twisted cases,
 since those higher energy levels are not going to be utilized to extract the scattering parameters in
 the following analysis anyway.
 \footnote{The additional operators in each particular channel
 helps to stabilize the lowest energy values in each irrep although
 the actual values of these higher states are not utilized.}
 As a result, altogether $9$ energy levels are kept for the scattering analysis
 in the following.

\begin{table*}

\begin{tabular}{|c||c|c|c|c|c|c|c|}
\hline
\hline
$\btheta$ & Irrep &  \multicolumn{2}{c|}{$\Delta E [t_{\min},t_{\max}](\mu=0.003)$}  &  \multicolumn{2}{c|}{$\Delta E [t_{\min},t_{\max}](\mu=0.006)$}  &  \multicolumn{2}{c|}{$\Delta E[t_{\min},t_{\max}] (\mu=0.008)$}\\

\hline
 &  &pmode0 &pmode1 &pmode0 &pmode1 &pmode0 &pmode1\\

\hline
$\bzero$
& $T_1$ &~0.001(2)~[8,13]~ &~0.068(3)~[6,11]~ &~0.005(2)~[9,14]~ &~0.059(4)~[8,13]~ &~0.003(1)~[8,13]~
&~0.046(3)~[8,13]~\\

\hline
\multirow{2}{*}{$(0, 0, \frac{\pi}{2})$}
&$A_1$ &0.001(3)~[8,15]  & &0.012(2)~[6,11]  & &0.007(2)~[8,13]  &\\
&$E$   &0.008(2)~[6,11]  & &0.013(1)~[6,11]  & &0.008(2)~[9,14] & \\

\hline
$(0, 0, \pi)$
&$E$   &0.004(5)~[10,15] & &0.019(2)~[8,13]  & &0.017(2)~[9,14] & \\

\hline
\multirow{3}{*}{$(\pi, \pi, 0)$}
&$B_1$  &0.027(3)~[8,15]   & &0.035(3)~[9,15]  & &0.027(1)~[7,12] & \\
&$B_2$  &0.032(3)~[8,13]   & &0.030(3)~[9,15]  & &0.025(2)~[8,14] & \\
&$B_3$  &0.032(2)~[7,13]   & &0.038(2)~[7,12]  & &0.028(2)~[9,14] & \\

\hline
$(\pi, \pi, \pi)$
&$E$ &0.040(2)~[5,10] & &0.047(1)~[5,10] & &0.038(3)~[7,12] & \\
\hline
\hline
\end{tabular}
\caption{Results for the energy shifts $\Delta E$ obtained in our calculations for
various cases. The time interval $[t_{\min},t_{\max}]$ from which we extract the values of $\Delta E$ are
also listed. These ranges are relevant for the estimation of the error for the zeta functions as
described in the text.\label{tab:deltaE}}
\end{table*}

\subsection{Extraction of scattering information}
 \label{subsec:scattering-parameter-fitting}

  The energy considered in this study is very close to the threshold of the
  $D^{*}$-$\bar{D}^{*}$ system, therefore one has the following effective range expansion:
 \begin{equation}
 \label{eq:kcotk-expansion}
  {k^{2l+1}\cot\delta_l (k)}=a^{-1}_l+{1\over2} r_{l} k^2 +\cdots~,
 \end{equation}
 where $a_{l}$ is the so-called scattering length for partial wave $l$ and $r_{l}$ is the
 corresponding effective force range while $\cdots$ represents terms that are higher order in $k^2$.
 It is more convenient to use a dimensionless form in our analysis.
 With $q^{2}=k^{2}L^{2}/(2\pi)^{2}$,
 Eq.~(\ref{eq:kcotk-expansion}) can be rewritten in terms of $q^2$:
  \begin{equation}
  {q^{2l+1}\cot\delta_l (q^2)}=B_l+{1\over2} R_{l} q^2 +\cdots~,
  \label{qcotangent}
 \end{equation}
 with $B_l=[L/(2\pi)]^{2l+1}a^{-1}_l$ and $R_l=[L/(2\pi)]^{2l-1}r_l$.
 In the following, we will call parameters $B_l$ and $R_l$ the low-energy scattering parameters in
 partial wave $l$ and our task is to extract these parameters from our simulation data.
 Since we have a definite lattice size and lattice spacing, it turns out that
 $q^2\simeq 1$ corresponds to $k^2\simeq [580{\rm MeV}]^2$ in physical units.

 It is also well-known that, close to the threshold, scattering is dominated by phase shifts
 coming from lower partial waves as long as they are non-vanishing. Therefore all partial waves
 with $l\ge 2$ will be ignored in the L\"uscher formula for this study.
 As mentioned in previous section, the irreps studied in this paper all preserve parity
 except for the case of $\btheta=(0,0,\pi/2)$ which breaks parity. Using the terminology in
 Ref.~\cite{Chen:2013omd}, this is the only parity-mixing scenario while all other points
 belong to the parity-conserving scenario.
 Thus to extract these low-energy scattering parameters from the lattice data,
 we have altogether $9$ points for different $q^2$ values: $2$ points in the parity-mixing case
 with $\btheta=(0,0,\pi/2)$ and $7$ points in the parity-conserving case.
 These are all tabulated in Table~\ref{tab:deltaE}.

 As all contributions from $l\ge 2$ partial waves have been neglected,
 the parity-conserving data (7 points) will depend only on the $s$-wave parameters $B_0$ and $R_0$
 while the parity-mixing data (2 points) will depend on both the $s$-wave parameters and
 the $p$-wave parameters $B_1$ and $R_1$. There are 3 different strategies to follow here:
 \begin{enumerate}
 \item A combined correlated fit using all $9$ data points. This yields the low-energy scattering parameters
 for both $s$-wave and $p$-wave;
 \item A correlated fit using only the parity-conserving points. This yields only the $s$-wave low-energy scattering
 parameters.
 \item A correlated fit using all data points, neglecting the parity-mixing effects
 of the two data points for $\btheta=(0,0,\pi/2)$. This also only yields the $s$-wave
 scattering parameters.
 \end{enumerate}
 We will first describe the fitting process following strategy 1 listed above.
 The other strategies follow similarly and the results will also be listed for
 comparisons.

 To be specific, in the parity-conserving case, we define
 \begin{equation}
 y_0(q^2)=m_{00}(q^2) ~.
 \end{equation}
 According to L\"uscher's formula Eq.~(\ref{eq:luscher_cube}), this should be equal to
 \begin{equation}
 q\cot\delta_0(q^2)=m_{00}(q^2)={1\over \pi^{3/2}}\calZ_{00}(1;q^2)~,
 \end{equation}
 for the non-twisted case while for the twisted case,
 one simply replace the corresponding zeta function by
 $\calZ^{\btheta}_{00}(1;q^2)$ \cite{Ozaki:2012ce}.
 In the parity-mixing case, however, things are more complicated. Apart from the
 $s$-wave phase shift $\delta_0(q^2)$, L\"uscher formula
 will also involve $\delta_1(q^2)$. Accordingly, we define
 \be
 y_1(q^2)=[m_{01}(q^2)]^2\;,
 \ee
 and, according to L\"uscher's formula, it is equal to
 \be
 y_1(q^2)=[q\cot\delta_0(q^2)-m_{00}][q\cot\delta_1(q^2)-m_{11}]
 \ee
 where the functions $m_{00}$, $m_{01}$ and $m_{11}$ are related
 to the corresponding zeta-functions, see e.g. Ref.~\cite{Ozaki:2012ce}.

 For definiteness, we label the data points as follows: the
 parity-conserving data points are labelled from $1$ to $N_0=7$
 while the parity-mixing points are labelled from $N_0+1=8$ to
 $N_0+N_1=7+2=9$. For later convenience, we also
 introduce an index function as follows,
 \be
 ind(I)=\left\{ \begin{aligned}
 0\;\; & \mbox{for $1\le I\le N_0$}\\
 1\;\; & \mbox{for $N_0+1\le I \le N_0+N_1$}
 \end{aligned}\right.
 \ee
 In other words, $ind(I)=0$ for the first $N_0$ parity-conserving data points
 while $ind(I)=1$ for the next $N_1$  parity-mixing data points. So our previous
 definitions of $y_0(q^2)$ and $y_1(q^2)$ may be written
 collectively as $y_{ind(I)}(q^2_I)$ with $I=1,2,\cdots, (N_0+N_1)$.
 We can then construct the $\chi^2$ function as usual
 \begin{equation}
 \begin{split}
 \label{eq:target_chi2}
 \chi^2=\sum^{N_0+N_1}_{I,J=1}
 &\left[F_{ind(I)}(q^2_I;\alpha)-y_{ind(I)}(q^2_I)\right]C^{-1}_{IJ}\\
 &\left[F_{ind(J)}(q^2_J;\alpha)-y_{ind(J)}(q^2_J)\right]
 \;.
 \end{split}
 \end{equation}
 where for $ind(I)=0,1$ the corresponding functions are
 (using the symbol $\alpha$ to collectively denote all
 the relevant fitting parameters $B_0$, $R_0$, $B_1$ and $R_1$):
 \ba
 &&\!\!\!\!\!\!F_0(q^2;\alpha)=B_0+{R_0\over 2}q^2\;,\\
 &&\!\!\!\!\!\!F_1(q^2;\alpha)=[B_0+{R_0\over 2} q^2-m_{00}]
 [B_1+{R_1\over 2} q^2-m_{11}].
 \ea
 For the estimation of the covariance matrix $C_{IJ}$ and also the errors
 for the zeta-functions that appear in the above formulas, we closely follow the steps
 outlined in Ref.~\cite{Chen:2013omd}. The reader is referred to that reference
 for further details. Basically, minimizing the target $\chi^2$ function in Eq.~(\ref{eq:target_chi2}),
 one could obtain all the parameters, namely $B_0$, $R_0$, $B_1$ and $R_1$,
 in a single step with all of our data.
  In the course of inverting the covariance matrix $C$ in Eq.~\ref{eq:target_chi2},
 the eigenvalues of the covariance matrix
 for each irrep have been calculated with both QR decomposition and singular value decomposition.
 The corresponding results show that the matrices are nonsingular.

  \begin{figure}[htb]
   {\resizebox{0.5\textwidth}{!}{\includegraphics{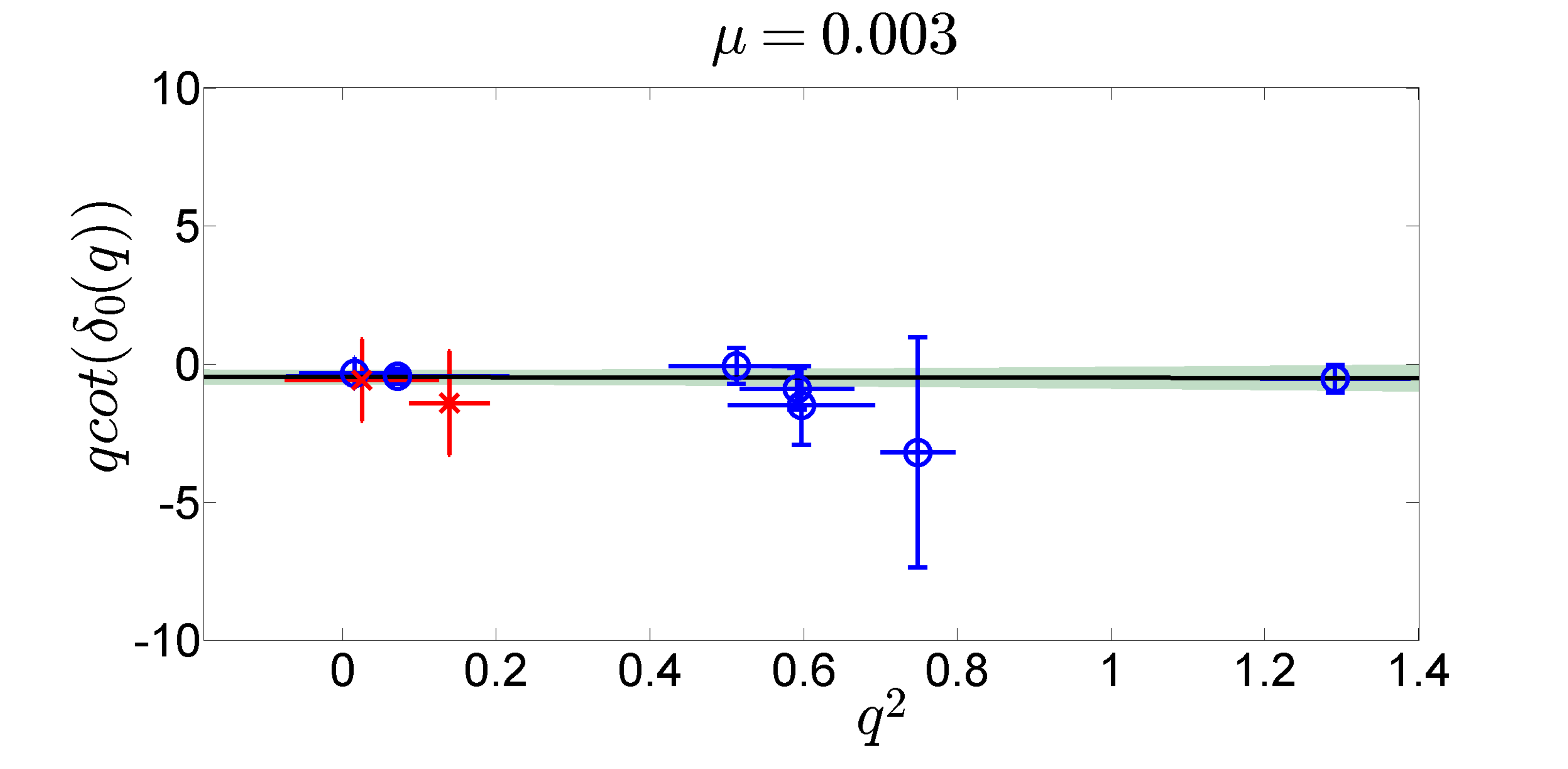}}}
    {\resizebox{0.5\textwidth}{!}{\includegraphics{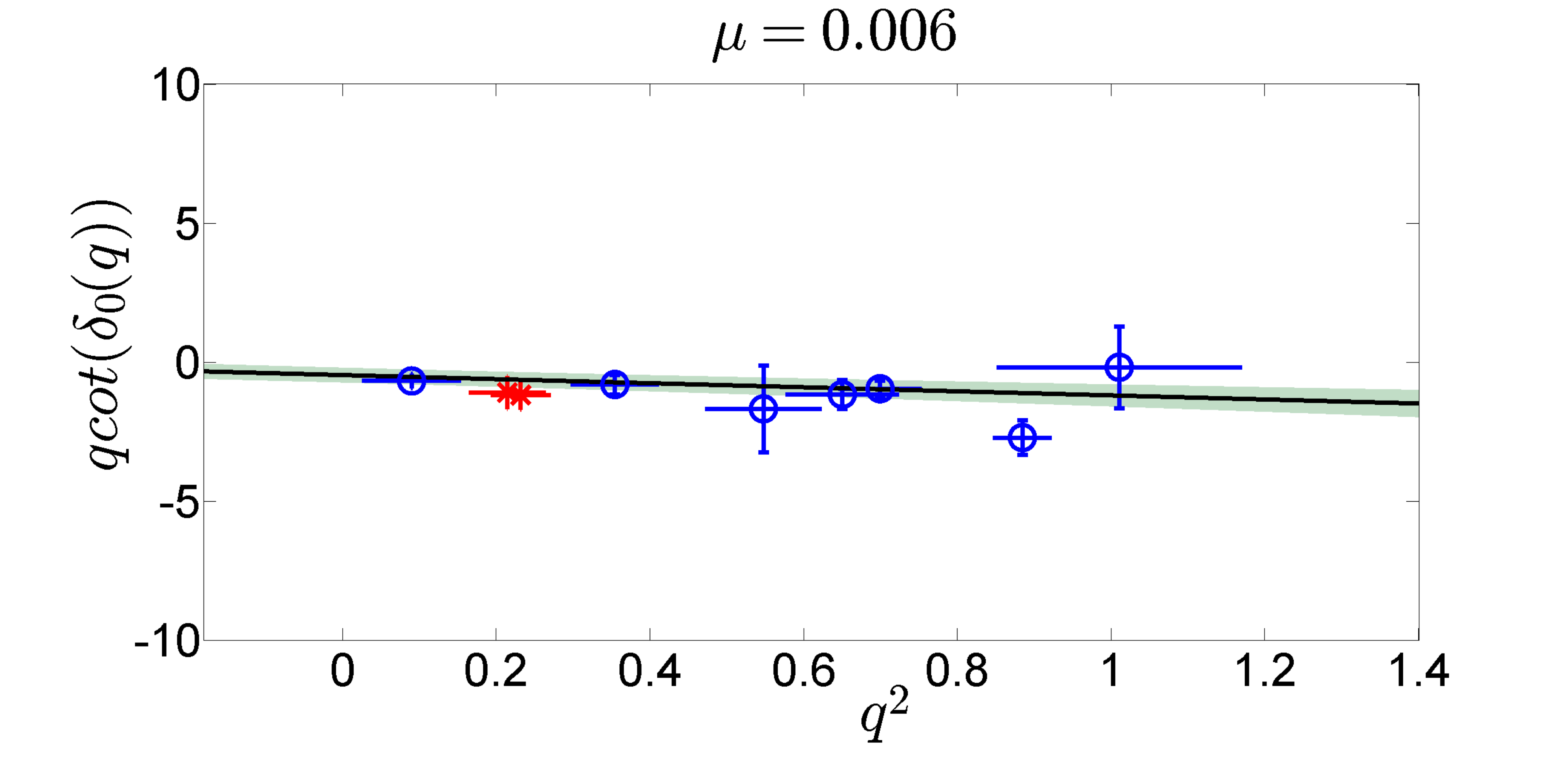}}}
     {\resizebox{0.5\textwidth}{!}{\includegraphics{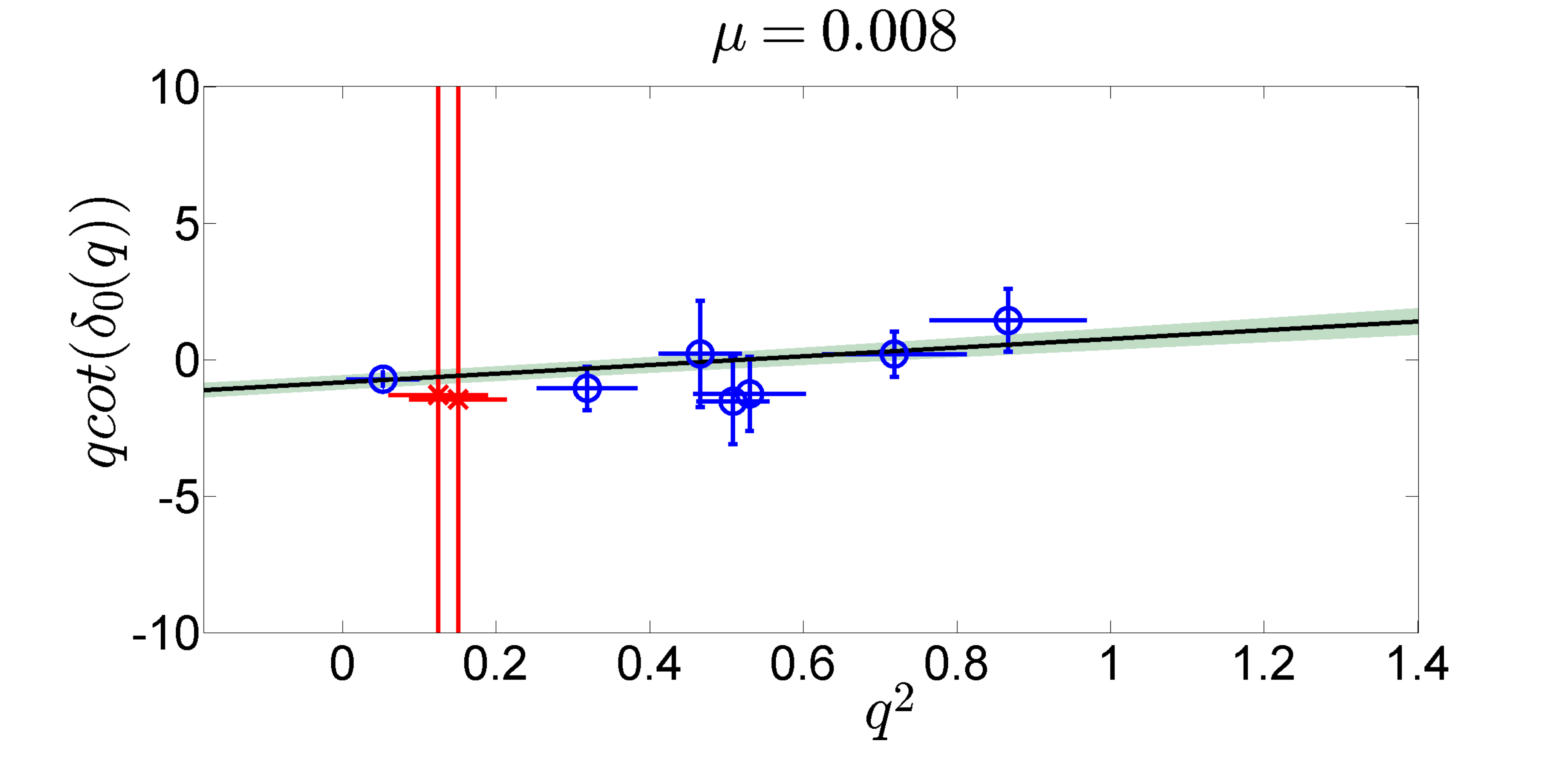}}}
 \caption{Results for the correlated fits from strategy 1 as described in the text. Each panel, from
 top to bottom, corresponds to $\mu=0.003, 0.006$ and $0.008$, respectively.
 The quantity $q\cot\delta_0(q^2)$ is plotted versus $q^2$ for all our data points,
 both parity-conserving case (blue open circles)
 and parity-mixing case (red crosses). The straight lines indicates the fitted
 result for $q\cot\delta_0(q^2)=B_0+(R_0/2)q^2$ and the shaded bands indicates the corresponding
 uncertainties. \label{fig:q2-qcotdeltaq-relation}}%
  \end{figure}
 To get a feeling of these fits, we plot
 the quantity $q\cot\delta_0(q^2)$ vs. $q^2$ in Fig.~\ref{fig:q2-qcotdeltaq-relation}
 obtained from strategy 1.
 The values of $q\cot\delta_{0}(q^2)$ for the data points are obtained via the relation
 \begin{equation}\label{eq:q2-m00-twochannel}
 q\cot\delta_{0}(q^2)=m_{00}+\frac{m^{2}_{01}}{q^{3}\cot\delta_{1}(q^2)-m_{11}}~,
 \end{equation}
 where the quantity $q^{3}\cot\delta_{1}(q^2)$ on the r.h.s of the above equation is replaced
 by $B_{1}+\frac{1}{2}R_{1}q^2$ with the fitted values for $B_{1}$ and $R_{1}$.
 This figure illustrates the situation for all three pion masses in our simulation.
 From top to bottom, each panel corresponds to $\mu=0.003$, $\mu=0.006$ and $\mu=0.008$, respectively.
 All data points obtained from our simulation are plotted in these figures.
 The blue open circles are the data points in the parity-conserving cases
 while the two red crosses in each panel are the data for the parity-mixing case.
 The straight lines in the figure illustrates
 the fitting function $F_0(q^2;\alpha)=B_0+(R_0/2)q^2$
 and the shaded bands indicate the corresponding uncertainties.
 As is seen from the figure, we do get a reasonable fit for all three
 pion mass values.
  Finally, the fitted values for the scattering parameters are
 summarized in Table~\ref{tab:fit-results-twochannel} for three values of $m^2_\pi$  in our simulation.

\begin{table}[htb]
\begin{tabular}{|c||c|c|c|c|c|}
\hline
\hline
  $\mu$ &$B_0$ &$R_0$ &$B_1$ &$R_1$  &$\chi^2/dof$ \\
  \hline
  0.003 &-0.47(35) & -0.051(243) &0.32(17) &-9.15(2.46) &1.66/5 \\
 \hline
  0.006  &-0.46(23) & -1.46(1.38) &-0.14(05) &-0.59(32) &9.92/5\\
  \hline
 0.008 & -0.83(17) & 3.18(2.12)  &0.85(31) &-14.42(5.83) &3.52/5\\
 \hline
 \hline
\end{tabular}
\caption{Fit results with strategy 1.\label{tab:fit-results-twochannel}}
\end{table}

\begin{table}[!htb]
\begin{tabular}{|c||c|c|c|}
\hline
\hline
 $~\mu~$  &$~B_0~$ &$~R_0~$  &$~\chi^2/dof~$ \\
  \hline
~0.003~ &~ -0.42(38)~ & ~-0.13(43)~  &~1.62/5~\\
 \hline
0.006  &-0.47(19) & -1.45(1.16)  &9.92/5 \\
  \hline
0.008 & -0.84(13) & 3.18(2.29)  &3.52/5\\
 \hline
 \hline
\end{tabular}
\caption{Fit results with strategy 2.\label{tab:fit-results-consevingpoint} }
\end{table}

\begin{table}[!htb]
\begin{tabular}{|c||c|c|c|}
\hline
\hline
  $\mu$ &$B_0$ &$R_0$  &$\chi^2/dof$ \\
  \hline
~0.003~  &~ -0.57(27)~ & ~0.15(59)~  &~2.44/7~\\
 \hline
0.006   &-0.33(21) & -1.72(1.25)  &10.94/7\\
  \hline
0.008  & -0.61(3) & 2.94(2.58)  &5.04/7 \\
 \hline
 \hline
\end{tabular}
\caption{Fit results with strategy 3.\label{tab:all-fit-results} }
\end{table}

 \begin{figure}[htb]
   {\resizebox{0.5\textwidth}{!}{\includegraphics{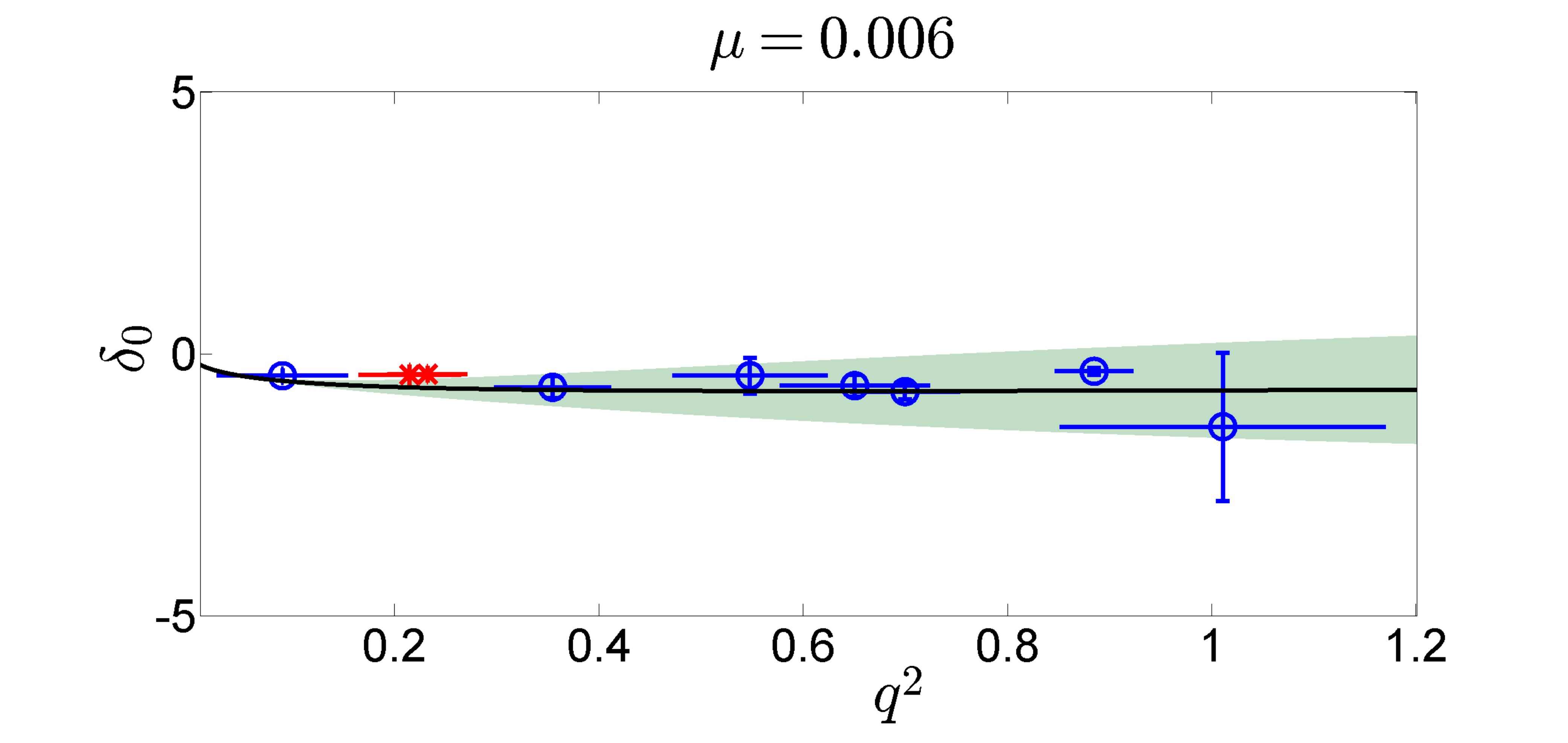}}}
 \caption{The same as Fig.~\ref{fig:q2-qcotdeltaq-relation}, but the comparison is done
 for the $s$-wave phase shift $\delta_0(q^2)$ itself. This is the case of $\mu=0.006$. \label{fig:phase-q2-result}}%
  \end{figure}
 To check the validity of the effective range expansion,
 we may also compare the $s$-wave phase shift $\delta_0(q^2)$ itself as
 a function $q^2$. The situation is shown in FIG.~\ref{fig:phase-q2-result}
 for $\mu=0.006$.

 Similarly, one could follow strategy 2 listed above and obtain
 the $s$-wave parameters using only the parity-conserving
 data points, i.e. the first $7$ data points. Or, alternatively
 following strategy 3 and obtain the $s$-wave scattering parameters
 using all the data points by neglecting the mixing between the
 $s$-wave and $p$-wave. Numerically this amounts to setting
 the matrix elements $m_{01}=0$ compared with the diagonal ones.
 The results obtained from these strategies can be compared with
 what we get from strategy 1. It turns out that, the mixing of the
 $s$- and $p$-wave indeed has little impact on the final results
 for the $s$-wave scattering parameters.
 The corresponding fitted results are summarized in
 Table~\ref{tab:fit-results-consevingpoint} and Table~\ref{tab:all-fit-results}, respectively.

 As is seen from these tables, as far as the $s$-wave scattering parameters are
 concerned, it seems that the parity-conserving data dominate the final fitting results.
 This is illustrated by consistent values for $B_0$ and $R_0$ in
 Table~\ref{tab:fit-results-twochannel} and Table~\ref{tab:fit-results-consevingpoint}.
 Finally, we regard our correlated fits from strategy 1 with all of our data as being more reliable
 and they are taken as our final results for this paper.

 \subsection{Physical values for the scattering parameters}
 \label{subsec:scattering-parameter-physical}

 The relation between the fitted values of $B_l$, $R_l$ for $l=0,1$
 and scattering parameters can be expressed as follows:
\begin{equation}
 a_l=\left({L\over 2\pi}\right)^{2l+1}\left({1\over B_l}\right)~,
 \;\;
 r_l=R_l\left({2\pi \over L}\right)^{2l-1}~.
 \end{equation}
 Taking the numbers of correlated fitting in Table~\ref{tab:fit-results-twochannel},
 for the $s$-wave, we can obtain the scattering length $a_0$: $-0.72(54)$fm, $-0.74(37)$fm,
 $-0.41(8)$fm for $\mu=0.003$, $0.006$, $0.008$, respectively.
 The values for $r_0$ can be also obtained accordingly.
 These numbers are summarized in Table~\ref{tab:result-physical-twochannel}.

 From strategy 2 and strategy 3, the correlated fitting are also conducted to obtain
 the scattering length $a_{0}$ and effective force range $r_{0}$. The specific results
 are summarized in Table~\ref{tab:result-physical-7point} and Table~\ref{tab:result-physical}, respectively.

  \renewcommand\arraystretch{1.2}
 \begin{table}[!htb]
\begin{tabular}{|c||c|c|c|}
\hline
\hline
  & $\mu=0.003$ &$\mu=0.006$  &$\mu=0.008$ \\
  \hline
  $a_0$[fm]~&~ -0.72(54)~ & ~-0.74(37)~    &~ -0.41(8)~\\
 \hline
  $r_0$[fm]& -0.018(83) & -0.50(43)    & 1.08(73)\\
 \hline
 \hline
\end{tabular}
\caption{The values for $a_0$ and $r_0$ in physical units obtained from
the numbers for the correlated fit in Table~\ref{tab:fit-results-twochannel},
from strategy 1. \label{tab:result-physical-twochannel}}
\end{table}

 \renewcommand\arraystretch{1.2}
 \begin{table}[!htb]
\begin{tabular}{|c||c|c|c|}
\hline
\hline
  & $\mu=0.003$ &$\mu=0.006$  &$\mu=0.008$ \\
  \hline
  $a_0$[fm]~&~ -0.80(71)~ & ~-0.73(30)~    &~ -0.41(6)~\\
 \hline
  $r_0$[fm]& -0.043(146) & -0.49(46)    & 1.09(78)\\
 \hline
 \hline
\end{tabular}
\caption{The values for $a_0$ and $r_0$ in physical units obtained from
the numbers for the correlated fit in Table~\ref{tab:fit-results-consevingpoint},
from strategy 2. \label{tab:result-physical-7point}}
\end{table}

\renewcommand\arraystretch{1.2}
 \begin{table}[!htb]
\begin{tabular}{|c||c|c|c|}
\hline
\hline
  & $\mu=0.003$ &$\mu=0.006$  &$\mu=0.008$ \\
  \hline
  $a_0$[fm]~&~ -0.60(28)~ & ~-1.03(65)~    &~ -0.56(3)~\\
 \hline
  $r_0$[fm]& 0.05(20) & -0.59(43)    & 1.00(88)\\
 \hline
 \hline
\end{tabular}
\caption{The values for $a_0$ and $r_0$ in physical units obtained from
the numbers for the correlated fit in Table~\ref{tab:all-fit-results},
from strategy 3. \label{tab:result-physical}}
\end{table}

 \subsection{Scattering parameters using the bootstrap method}
 \label{subsec:bootstrap}

 The errors used in the analysis discussed so far are
 estimated using the jackknife method.
 To crosscheck these results, bootstrap method is also
 utilized to analyze directly the final scattering length $a_{0}$ and effective range $r_{0}$.
 The specific procedure is as follows.
 \begin{enumerate}
 \item Select randomly 200 configurations from the given configurations in Table~\ref{tab:parameter} for
 each parameter $\mu$. Do the selection $N_{random}$ times and label each sample by an
 integer $i$, $i=1,2,\cdots, N_{random}$.
 \item For each randomly selected sample $i$,
 repeat the analysis process described so far in section~\ref{sec:simulation-details}.
 This yields one set of scattering parameters, say $1/a^{i}_{0}$ and $r^{i}_{0}$.
 \item Analyze the distribution of these values.
 Taking $r_{0}$ as an example, find the values $p$ and $q$ so that these bracket
 the central 68\% of the  $r^{i}_{0}$ values:
 \begin{equation}
 \frac{N(r^{i}_{0}<p)}{N_{random}}=0.16 ~~~~~~~\frac{N(r^{i}_{0}>q)}{N_{random}}=0.16
 \end{equation}
 where $N(r_{0}<p)$ denotes the number of $r^{i}_{0}$ satisfying $r^{i}_{0}<p$.
 \item the bootstrap estimate of the asymmetric errors for the quantity can be given as:
 \begin{equation}
 r_{0} = \langle r_{0}\rangle^{+(q-\langle r_{0}\rangle)}_{-(\langle r_{0}\rangle-p)}
 \end{equation}
 with $\langle r_{0}\rangle$ denoting the weighted mean of $\{r^{i}_{0}\}$.
 \end{enumerate}

\begin{figure}[!htb]
   {\resizebox{0.5\textwidth}{!}{\includegraphics{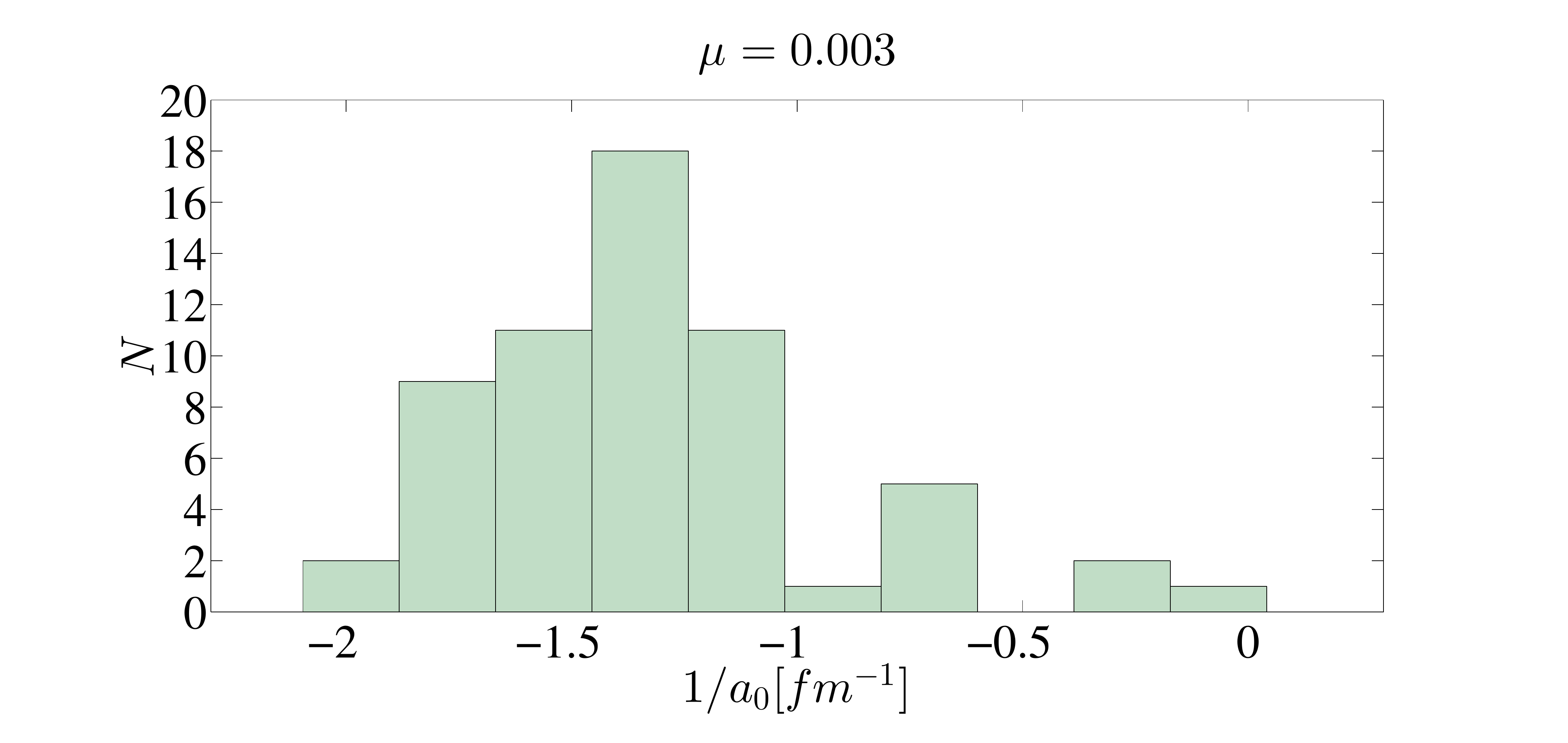}}}
   {\resizebox{0.5\textwidth}{!}{\includegraphics{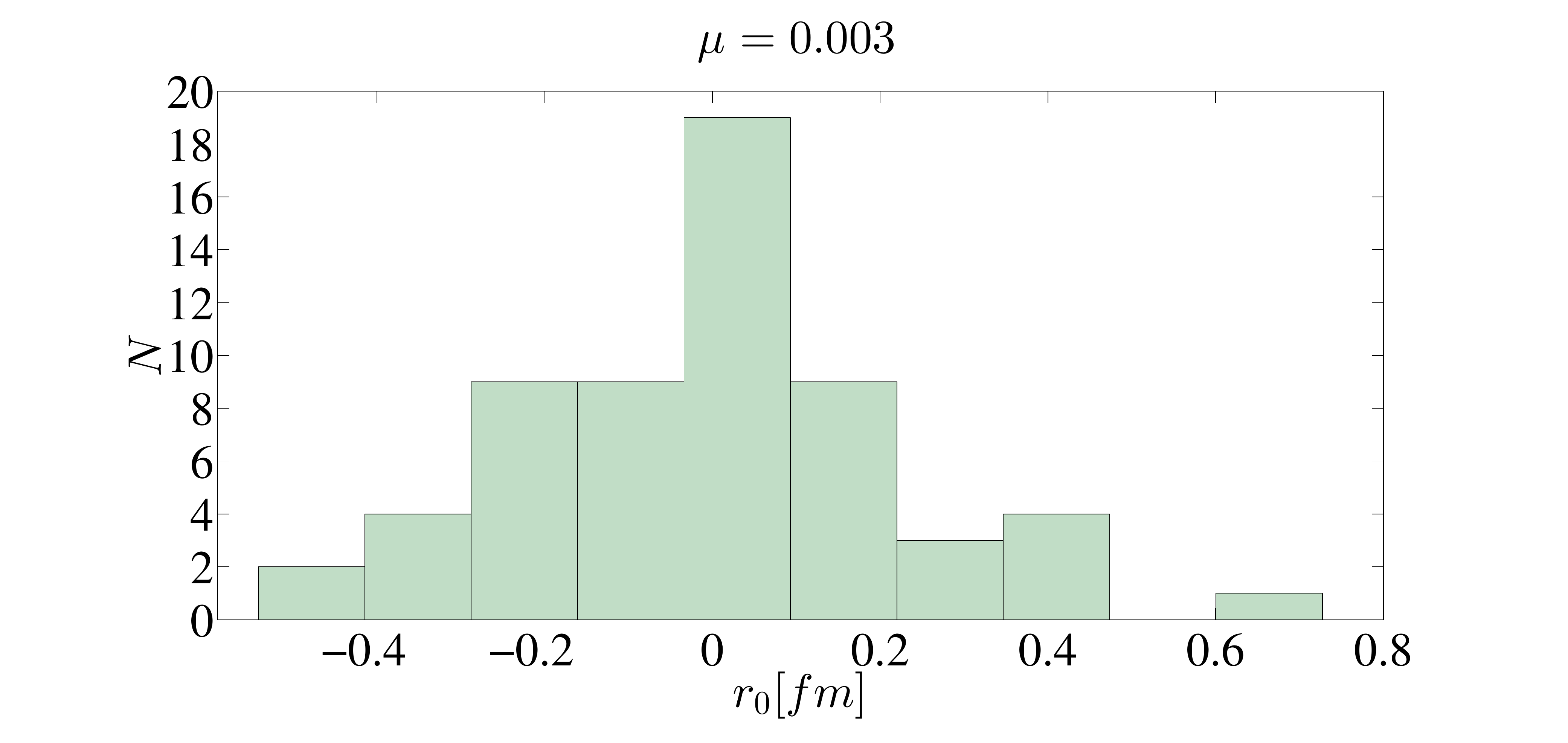}}}
   \caption{Distribution of \{$\frac{1}{a^{i}_{0}}$\} and \{$r^{i}_{0}$\}
   from strategy 3 for bootstrap method at $\mu=0.003$. \label{fig:bootstrap-recia0-r0-mu003}}%
  \end{figure}

 In this work, we take $N_{random}=60$ at three different $\mu$ ($\mu=0.003,0.006,0.008$)
 to estimate the bootstrap error of scattering parameters with strategy 2 described in section IV.
 As an illustration, the distribution of $1/a^{i}_{0}$ and $r^{i}_{0}$
 for case $\mu=0.003$ are shown in FIG.~\ref{fig:bootstrap-recia0-r0-mu003}.
 Meanwhile, the asymmetric error of $a_{0}$ and $r_{0}$ are estimated with these
 $60$ samples. These final specific values are summarized
 in Table~\ref{tab:a0-r0-results-bootstrap-7point}.

 Alternatively, we have also estimated the bootstrap error of scattering parameters with
 strategy 3. The only difference is that the data points corresponding to
 $A_{1}$ and $E$ irreps $(\btheta=(0,0, \pi/2))$ are not left out during the process, that is
 neglecting the parity-mixing of effects of the two data points.
 The final results are summarized in Table~\ref{tab:a0-r0-results-bootstrap}.

\renewcommand\arraystretch{1.5}
 \begin{table}[!htb]
 \begin{tabular}{|c|c|c|c|}
 \hline
 \hline
  & $\mu=0.003$ &$\mu=0.006$  &$\mu=0.008$ \\
  \hline
  $a_{0}$[fm]~&~ $-0.74^{+0.21}_{-0.15}$~ & ~$-0.68^{+0.19}_{-0.18}$~  &~$-0.48^{+0.22}_{-0.22}$~\\
 \hline
  $r_{0}$[fm] &$-0.0048^{+0.18}_{-0.20}$  & $-0.038^{+0.12}_{-0.16}$    &~$0.73^{+0.58}_{-0.62}$\\
 \hline
 \hline
 \end{tabular}
 \caption{The values for $a_{0}$ and $r_{0}$ from strategy 2 with parity-conserving data points
 with bootstrap method at $\mu=0.003,0.006,0.008$.\label{tab:a0-r0-results-bootstrap-7point}}
 \end{table}

\renewcommand\arraystretch{1.5}
 \begin{table}[!htb]
 \begin{tabular}{|c|c|c|c|}
 \hline
 \hline
  & $\mu=0.003$ &$\mu=0.006$  &$\mu=0.008$ \\
  \hline
  $a_{0}$[fm]~&~ $-0.76^{+0.14}_{-0.21}$~ & ~$-0.86^{+0.22}_{-0.22}$~  &~$-0.59^{+0.19}_{-0.25}$~\\
 \hline
  $r_{0}$[fm] &$-0.0022^{+0.18}_{-0.19}$  & $-0.14^{+0.15}_{-0.18}$    &~$0.64^{+0.50}_{-0.51}$\\
 \hline
 \hline
 \end{tabular}
 \caption{The values for $a_{0}$ and $r_{0}$ from strategy 3 with all data points with bootstrap method
 at $\mu=0.003,0.006,0.008$.\label{tab:a0-r0-results-bootstrap}}
 \end{table}
 As observed from Table~\ref{tab:result-physical}, Table~\ref{tab:result-physical-7point},
 Table~\ref{tab:a0-r0-results-bootstrap},
 and Table~\ref{tab:a0-r0-results-bootstrap-7point},
 jackknife method and bootstrap method yield compatible results.

 \subsection{Implication of our results}
 \label{subsec:implications}

 As is said, we take the fitted result from strategy 1 as our final results,
 i.e. those in Table~\ref{tab:fit-results-twochannel} and Table~\ref{tab:result-physical-twochannel}.
 Based on our results, the values of $a_0$ do not seem to follow a regular chiral extrapolation pattern,
 at least not within the range that we have studied. We therefore kept the individual values
 for $a_0$ and $r_0$ for each case. This irregularity might be caused by the smallness
 of the value $m_\pi L\sim 3.3$ for $\mu=0.003$.
 To circumvent this, one has to study a larger lattice.

 The negative values of the parameter $B_0$ (hence the scattering length $a_0$) indicates
 that the two constituent mesons for the $(D^{*}\bar{D}^*)^\pm$ system
 have weak repulsive interactions at low energies.
 Therefore, our result does not support the bound state scenario for these two mesons.

 Another check for the possible bound state would be
 to look for those negative $q^2$ values we obtained which corresponds to
 the negative values of $\delta E$ listed in Table~\ref{tab:deltaE}.
 However, the $q^2$ for different channel in our study are all positive which
 contradicts the possibility of a bound state.
 Since the cases we are studying is still far from the physical pion mass case,
 we still cannot rule out the possibility the appearance of a bound state once
 the pion mass is lowered (and the lattice size $L$ is also increased accordingly
 to control the finite volume corrections).
 Such scenarios do occur in lattice studies of two nucleons.

 \section{Conclusions}
  \label{sec:conclusions}

 In this paper, the low-energy scattering of $D^{*}$ and $\bar{D}^{*}$
 is studied with $N_{f}=2$ twisted mass fermion configurations.
 In our calculation, three different pion mass values ($m_{\pi}=300,420,485$~MeV)
 are utilized to investigate the pion mass dependence,
 and the corresponding lattice size is $32^{3}\times 64$ with
 a lattice spacing $a\simeq 0.067$~fm. We have used twisted boundary
 conditions to enhance the momentum resolution close to the
 threshold. Using L\"uscher's finite-size technique, the $s$-wave scattering
 in the channel $J^{P}=1^{+}$ is studied and the scattering
 parameters are obtained by correlated fitting procedure.
 As a crosscheck, two different statistical error estimating
 methods, jackknife and bootstrap method
 are utilized which yield compatible results for the $s$-wave scattering parameters.
 The results from a correlated fit with all of the data with errors estimated using
 the jackknife method is regarded as the final result for this paper.

 Our results indicate that, for all three pion mass values that we simulated,
  the scattering lengths are negative which indicates
  a weak repulsive interaction between the the two mesons
 ($D^{*}$ and $\bar{D}^*$ or its conjugated systems under $C$-parity or $G$-parity).
 Thus a bound state of the two mesons in $J^{P}=1^{+}$ channel
 is not supported based on our current lattice results.
 However, as we pointed out already, we cannot rule out the possibility of
 a bound state for the two vector charmed mesons when the pion mass is lowered and the volume is
 increased accordingly.  This requires further systematic lattice studies.
 Furthermore, it is also possible that more complete set of interpolation operators
 and a coupled channel study is required. In summary, this lattice study has shed
 some light on the nature of $Z^\pm_c(4025)$ however
 it remains to be clarified by future more systematic studies.

\section*{ACKNOWLEDGEMENTS}
  The authors would like to thank F.~K. Guo, B.~Knippschild, L.~Liu, U.~Meissner, A.~Rusetsky,
  and C.~Urbach for helpful discussions.
 The authors would like to thank the European Twisted Mass Collaboration (ETMC)
 to allow us to use their gauge field configurations. Our thanks also go to
 National Supercomputing Center  in Tianjin (NSCC) and the Bejing Computing Center (BCC) where part of the numerical computations are performed.
 This work is supported in part by the
 National Science Foundation of China (NSFC) under the project
 No.11335001,
 No.11275169,
 No.11075167,
 No.11105153.
 It is also supported in part by the DFG and the NSFC (No.11261130311) through funds
 provided to the Sino-Germen CRC 110 ``Symmetries and the Emergence
 of Structure in QCD''.
 M.~Gong and Z.~Liu are partially supported
 by the Youth Innovation Promotion Association of CAS (2013013, 2011013).


%

 \end{document}